\documentclass[11pt,aps,prc,
letter,preprintnumbers,tightenlines,
superscriptaddress,
nofootinbib]{revtex4}

\usepackage[T1]{fontenc}
\usepackage{lmodern}
\usepackage{charter}
\usepackage[expert]{mathdesign}

\usepackage{graphicx}
\usepackage{calc}
\usepackage{epsfig}
\usepackage{amsmath,amsfonts}
\usepackage{mathrsfs}
\usepackage{xcolor}
\usepackage{slashed}
\usepackage[colorlinks,citecolor=blue,linktoc=all,linkcolor=cyan,urlcolor=blue]{hyperref}

\def\beq{\begin{equation}}
\def\eeq{\end{equation}}
\def\bea{\begin{eqnarray}}
\def\eea{\end{eqnarray}}
\def\beqa{\begin{equation}\begin{array}{l}}
\def\eeqa{\end{array}\end{equation}}
\def\eqlab#1{\label{eq:#1}}

\def\Eqref#1{Eq.~(\ref{eq:#1})}

\def\half{\mbox{\small{$\frac{1}{2}$}}}

\def\barr{\left(\begin{array}{c}}
\def\earr{\end{array}\right)}
\def\bmat{\left(\begin{array}{cc}}
\def\emat{\end{array}\right)}
\def\al{\alpha}
\def\be{\beta}
\def\ga{\gamma} 
\def\de{\delta} \def\De{\Delta}\def\vDe{\varDelta}
  \def\eps{\epsilon}

 \def\La{{\Lambda}}

\def\si{\sigma}

\def\pa{\partial}

\def\pa{\partial}

\def\3d{3-D}

\def\AnswerYes{y}
\def\ShowLabelsVersion{n}         
\def\ShowChangesVersion{n}        
\def\ShowAnnotationsVersion{n}    
\ifx\ShowLabelsVersion\AnswerYes
 \usepackage{showkeys}
\fi
\ifx\ShowAnnotationsVersion\AnswerYes
  \newcommand{\comment}[1]{{\color{blue}\textit{#1}}}
  \newcommand{\margin}[1]{\marginpar{\scriptsize\sffamily\bfseries{#1}}}
\else
  \newcommand{\comment}[1]{}
  \newcommand{\margin}[1]{}
  
\fi
\ifx\ShowChangesVersion\AnswerYes
  \usepackage{ulem}
   
  \newcommand{\delete}[1]{\sout{#1}}            
  \renewcommand{\emph}[1]{\textit{#1}}           
\else

  \newcommand{\sout}[1]{}
  \newcommand{\xout}[1]{}

  \newcommand{\delete}[1]{}
\fi

\begin{document}
\preprint{MITP/16-118}
\title{Generalized polarizabilities of the nucleon  in baryon chiral 
perturbation theory}

\author{Vadim Lensky}\email{lensky@itep.ru}
\affiliation{Institut f\"ur Kernphysik, Cluster of Excellence PRISMA, Johannes Gutenberg Universit\"at Mainz,  D-55128 Mainz, Germany}
\affiliation{Institute for Theoretical and Experimental Physics, 117218 Moscow, Russia}
\affiliation{National Research Nuclear University MEPhI (Moscow Engineering Physics Institute), 115409 Moscow, Russia}

\author{Vladimir Pascalutsa}
\affiliation{Institut f\"ur Kernphysik, Cluster of Excellence PRISMA, Johannes Gutenberg Universit\"at Mainz,  D-55128 Mainz, Germany}

\author{Marc Vanderhaeghen}
\affiliation{Institut f\"ur Kernphysik, Cluster of Excellence PRISMA, Johannes Gutenberg Universit\"at Mainz,  D-55128 Mainz, Germany}
\date{\today}

\begin{abstract}
The nucleon generalized polarizabilities
(GPs), probed in virtual Compton scattering (VCS),
describe the spatial distribution of the polarization density in a nucleon. 
They are accessed 
experimentally via the process of electron-proton bremsstrahlung ($ep\to ep\gamma$) at electron-beam facilities,
such as MIT-Bates, CEBAF (Jefferson Lab), and MAMI (Mainz). 
We present the calculation of the nucleon GPs 
and VCS observables at
next-to-leading order in baryon chiral perturbation theory
(B$\chi$PT), and confront the results with the empirical information.
At this order our results are predictions, in the sense that all
the parameters are well-known from elsewhere. 
Within the relatively large uncertainties of our calculation
we find good agreement with the experimental observations of
VCS and the empirical extractions of the GPs.
We find large discrepancies with previous chiral
calculations---all done in heavy-baryon $\chi$PT (HB$\chi$PT)---and discuss the differences between B$\chi$PT and HB$\chi$PT
responsible for  these discrepancies. 
\end{abstract}


\pacs{ }
\date{\today}
\maketitle
\newpage
\tableofcontents

\section{Introduction}
\begin{figure}[tbh]
\includegraphics[width=\textwidth]{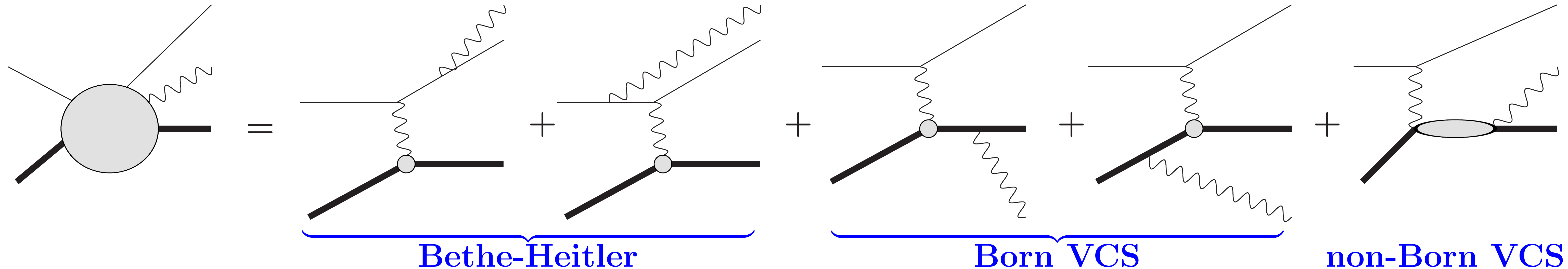}
\caption{Mechanisms contributing to $ep\to ep\gamma $
in the one-photon-exchange approximation:
Bethe-Heitler, Born VCS, non-Born VCS. Thick (thin) solid lines
denote the proton (the electron), wavy lines denote photons. Small circles
denote the interaction vertex of a proton with a virtual photon, and the ellipse
stands for the generic non-Born VCS amplitude.
}
\label{fig:Born_BH}
\end{figure}

Long after the early studies of the electron-proton ($ep$) bremsstrahlung \cite{Berg58,Berg61}, it was realized that this process holds the key
to the {\it generalized polarizabilities} (GPs) of the nucleon~\cite{Guichon:1995pu}; 
see Ref.~\cite{Guichon:1998xv} for a review. The GPs extend the concept of
static polarizabilities to finite momentum-transfer $Q^2$, and
have an interpretation of the distribution of polarization densities
in the nucleon \cite{Gorchtein:2009qq}. They naturally arise
in virtual Compton scattering (VCS) with the incoming virtual 
photon of spacelike virtuality
$Q^2$, and the outgoing real photon of very low frequency;
hence, the $ep$ bremsstrahlung
which, in the one-photon-exchange
approximation,  decomposes 
into the Bethe-Heitler (BH) process and VCS, cf.~Fig.~\ref{fig:Born_BH}.
Shown in the figure is also the split of 
VCS into: A) the {\it Born contribution} 
to VCS,
with the intermediate state being the nucleon itself, and 
B) {\it non-Born contribution} to VCS, 
which at low energies is entirely determined by GPs~\cite{Guichon:1995pu,Drechsel:1997xv,Drechsel:1998zm}.
The BH and Born VCS contributions
are given in terms of the electromagnetic nucleon form factors known from elastic electron scattering.
The non-Born VCS amplitude, carrying the
information about the ``inelastic'' structure of the nucleon, is the unknown piece that one is trying to
access in the $ep$ bremsstrahlung.

Over the past two decades the experimental studies clearly demonstrated 
the feasibility of an accurate extraction of 
proton GPs from $ep$ 
bremsstrahlung~\cite{Roche:2000ng,Bourgeois:2006js,Bourgeois:2011zz,d'Hose:2006xz,
Janssens:2008qe,Doria:2015dyx,Correa:thesis}. This experimental progress
has been echoed by theory advances.
 A number of impressive calculations have been done in heavy-baryon chiral perturbation
 theory (HB$\chi$PT)~\cite{Hemmert:1996gr,Hemmert:1997at, Hemmert:1999pz,Kao:2002cn,Kao:2004us}, albeit
 showing a rather poor convergence. A much more empirically viable theory of proton 
 GPs and VCS was developed by Pasquini
 {\it et al.}~\cite{Drechsel:2002ar,Pasquini:2001yy} 
 based on fixed-$t$ dispersive relations (DRs) for the VCS amplitudes.  
 Incidentally, this framework is used in many experimental studies
 to extract the GPs from the VCS observables.

The present work is aiming to advance the  chiral effective-field theoretic
approach by applying the manifestly Lorentz-invariant variant of baryon 
chiral perturbation theory (B$\chi$PT) to nucleon VCS and GPs. As many recent calculations  demonstrate (see, e.g., 
\cite{Becher:1999he,Fuchs:2003qc,Pascalutsa:2004ga,Holstein:2005db,Pascalutsa:2005vq,Ledwig:2011cx,Alarcon:2011zs,Alarcon:2012kn,Yao:2016vbz,Blin:2016itn}), 
B$\chi$PT shows an improved convergence over the analogous HB$\chi$PT calculations, and, as result, a more ``natural'' description of the nucleon polarizabilities and  Compton scattering processes \cite{Pascalutsa:2004wm,Hall:2012iw,Bernard:2012hb,Lensky:2009uv,Lensky:2014dda}. In this paper, we extend the previous B$\chi$PT calculations
of Lensky {\it et al.}~\cite{Lensky:2009uv,Lensky:2014dda,Lensky:2015awa}, 
done for nucleon polarizabilities appearing in real and forward 
doubly-virtual Compton scattering (RCS and 
VVCS, respectively),  to the case of GPs and VCS. As in the previous cases, the
present
calculation is ``predictive'' in the sense that it has no free parameters 
to be fixed by the empirical information from Compton processes. And, as in other cases, we
find significant improvements in convergence over the analogous HB$\chi$PT results. Arguably, the
main improvement is that our postdictions compare well with the experimental data
on VCS observables, at least given the significant theoretical uncertainties.

The paper is organized as follows. In Sec.~\ref{sec:VCS}, we open
with the general remarks concerning the connection between polarizabilities
and low-energy Compton scattering processes, and then focus on defining
the GPs and the VCS observables. 
Sec.~\ref{sec:BChPT} contains the details of our B$\chi$PT calculation, including
power-counting, diagrams, theory error estimate, and remarks on a number of technical 
issues which arise in these calculations. 
Sec.~\ref{sec:comparison} compares our calculation with previous estimates: the linear $\sigma$-model, 
HB$\chi$PT calculations, and fixed-$t$ dispersive estimates. 
Sec.~\ref{sec:results} confronts the results with the available experimental data.  
Sec.~\ref{sec:conclusions} contains the concluding remarks.
Appendix~\ref{sec:appendix:tensors} contains expressions for the tensors
that are used in the decomposition of the VCS amplitude, whereas 
Appendix~\ref{sec:appendix:amplitudes} contains analytic expressions
for those combinations of the invariant VCS amplitudes that contribute
to the GPs.

\section{Polarizabilities in Compton processes}\label{sec:VCS}

\begin{figure}[t]
\includegraphics[width=0.4\textwidth]{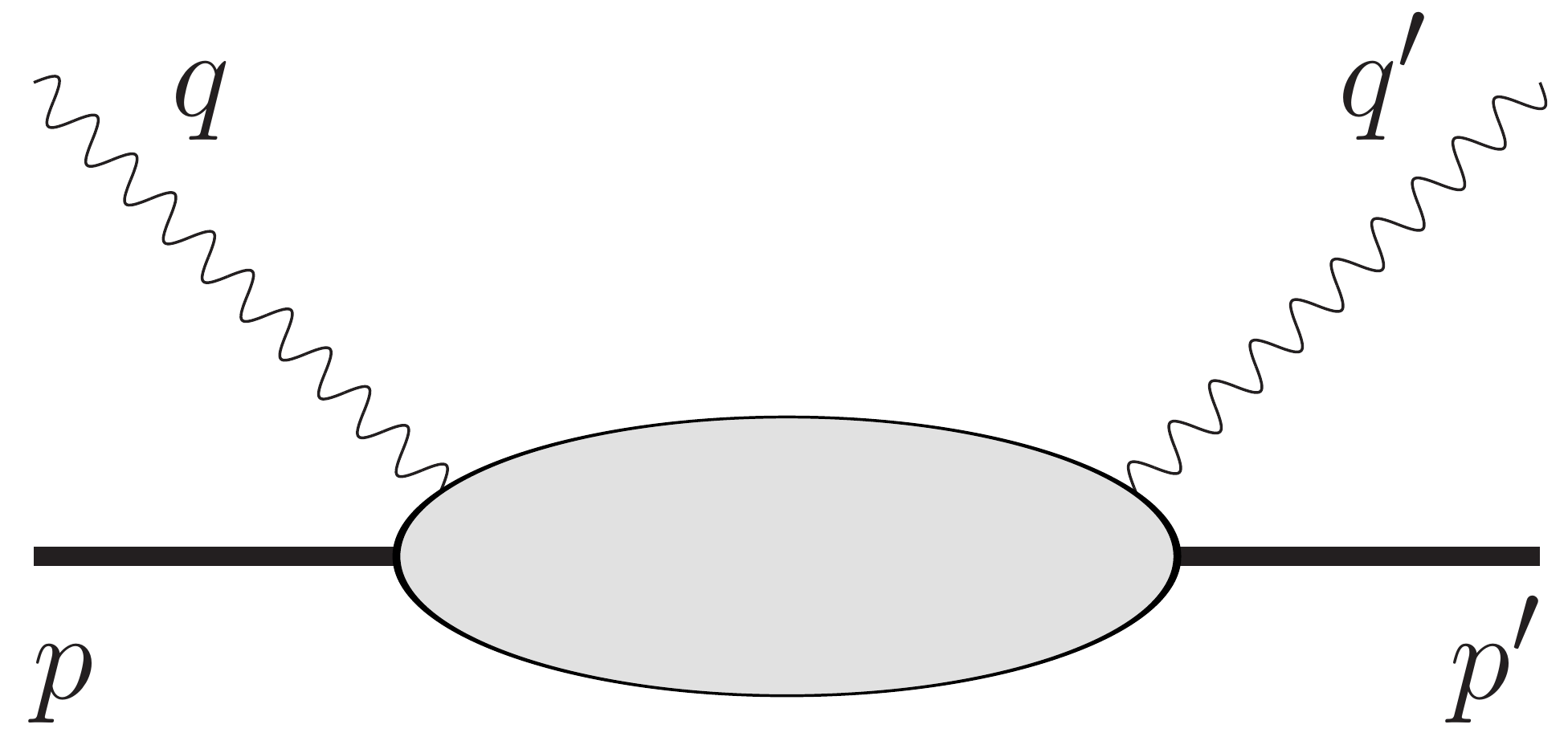}
\caption{General Compton scattering process. The notation is as in Fig.~\ref{fig:Born_BH}, with the four-momenta of the corresponding
particles denoted in the figure.
}
\label{fig:VCS_VCS}
\end{figure}

Let us start by pointing out that there are two different ways
of introducing the momentum-transfer dependence of polarizabilities:
one via the forward doubly-virtual Compton scattering (VVCS), 
the other via the single-virtual Compton scattering (VCS). 
To see the difference, consider a general Compton scattering (CS) process in Fig.~\ref{fig:VCS_VCS}, 
described by a number of scalar
amplitudes $A_i$, functions of Mandelstam invariants 
\beq 
s=(p+q)^2=(p'+q')^2, \quad t=(q-q')^2=(p'-p)^2,\quad  
u=(p-q')^2=(p'-q)^2.
\eeq 
The latter satisfy the usual kinematical constraint, 
\beq
s+u+t=2M^2+q^2+q^{\prime\, 2} ,
\eeq
with $M$ the nucleon mass and $q^2$, $q^{\prime\, 2}$ the photon virtualities. 
The polarizabilities can be equated 
with the coefficients in the low-energy  expansion of the CS amplitudes $A_i$. 
Introducing the invariant energies of the incoming and outgoing photon:
$\nu =p\cdot q/M$ and $\nu' =p\cdot q'/M$, the low-energy expansion requires
at least one of them to be small. Note that the kinematical constraint can be written as:
$t=-2M\, (\nu-\nu')$, hence by energy conservation $t\leq 0$. It is convenient
to introduce the kinematic invariant,
\beq 
\xi = \frac{s-u}{4M},
\eeq
which is odd under the photon crossing, $\xi\to -\xi$, whereas $t$ is even.
In the most general situation, the CS amplitudes are 
functions of four independent variables, e.g., the photon energies and virtualities,
$A_i = A_i(\nu,\nu'; q^2, q^{\prime\, 2})$, or equivalently, 
$A_i(\xi, t; q^2, q^{\prime\, 2})$.

Now, for real photons ($q^2=q^{\prime\, 2}=0$), small $\nu$ infers the smallness of $\nu'$ and $t$.
This is the limit in which the static polarizabilities are defined. 
For both photons virtual and having the same momentum $q'=q$, we
deal with the forward VVCS process which, by means of unitarity and causality, 
can be expressed in terms of nucleon structure functions. In this case the low-energy 
expansion of the non-Born amplitude\footnote{The Born amplitude will be treated exactly, as the convergence of its low-energy expansion is severely 
limited by the nucleon pole at $\xi=\pm Q^2/2M$.}
is around $\nu=\nu'=0$ and the polarizabilities arise as moments of structure functions.

In this work we are concerned with VCS---the Compton process 
where the initial photon is virtual ($q^2=-Q^2$)
and the final one is real ($q^{\prime\, 2}=0$). 
The polarizabilities are obtained by expanding around $\nu'=0$, while 
$\nu$ is required to be near the lowest physical value (i.e., the elastic threshold): $\nu_0=Q^2/2M$.
This limit corresponds with $t=-Q^2$. Note that our $Q^2$
corresponds to $\tilde{Q}^2$ in the notation of Ref.~\cite{Guichon:1995pu},
and to $Q_0^2$ in the notation of Ref.~\cite{Drechsel:1997xv}.
The three-momentum squared of the initial photon is given in this limit
by $\bar{\rm q}^2=Q^2(1+\tau)$, with $\tau=Q^2/4M^2=\nu_0/2M$.

To distinguish between the GPs arising in VVCS and VCS, we shall refer to the former ones as ``symmetric'' and to the latter ones as ``skewed''. The three considered situations can thus be classified as follows:
\begin{description}
\item[static (RCS):] $q^2=0$, $q^{\prime\, 2}=0$, $\nu\approx 0$, $\nu'\approx 0$, $t\approx 0$.
\item[symmetric GPs (VVCS):] $q^2=q^{\prime\, 2}=-Q^2$, $\nu=\nu'\approx 0$, $t=0$.
\item[skewed GPs (VCS):] $q^2=-Q^2$, $q^{\prime\, 2}=0$, $\nu'\approx 0$, $\nu\approx Q^2/2M$, $t\approx -Q^2$.
\end{description}
A nice pictorial representation of these situations can be found in Ref.~\cite{Eichmann:2012mp}.
Because the low-energy expansions in VVCS and VCS are performed around such different kinematical points, 
the symmetric and skewed GPs  are only connected in the static (real-photon) limit.

Let us now consider the low-energy expansion of VCS in more detail.
The low-energy theorem for VCS~\cite{Guichon:1995pu,Low:1954kd,Scherer:1996ux}
states that the expansion of the non-Born amplitude in powers of the final photon's
energy $\nu'$ starts at $O(\nu^{\prime\, 1})$, whereas the two leading
terms, $O(\nu^{\prime\,-1})$ and $O(\nu^{\prime\, 0})$, are entirely determined by the
tree-level BH and Born amplitudes. 
The leading-order [$O(\nu^{\prime\,1})$] non-Born
contribution had initially been parametrized by ten skewed GPs~\cite{Guichon:1995pu}. Soon after, it was discovered that 
the crossing symmetry [see Eq.~(\ref{eq:crossing})]
reduces the number of independent GPs to six~\cite{Drechsel:1997xv}. These six GPs are often denoted as\footnote{Originally \cite{Guichon:1995pu} they were denoted as, respectively: $P^{(01,01)0}$, $P^{(11,11)0}$, $P^{(01,01)1}$, $P^{(11,11)1}$, $P^{(11,02)1}$, $P^{(01,12)1}$.}
\begin{align}
P^{(L1,L1)0}(Q^2),\ P^{(M1,M1)0}(Q^2),\ P^{(L1,L1)1}(Q^2),\ P^{(M1,M1)1}(Q^2),\
P^{(M1,L2)1}(Q^2),\ P^{(L1,M2)1}(Q^2)\,,
\label{eq:GPs}
\end{align}
where $P^{(\rho'\ell',\rho\ell)S}(Q^2)$ correspond with 
a multipole amplitude (at $\nu'=0$) where $\rho =L, M$ denotes whether the photon is of the longitudinal or the magnetic type and $\ell$ 
denotes the angular momentum
(respectively, $\rho'\ell'$ or $\rho\ell$ for the final or initial photon); $S=1$ or $S=0$ indicates whether the transition involves the proton's spin flip or not.

To be more specific, we consider the tensor decomposition of the VCS amplitude $\mathcal{M}^{\mu\nu}$
into the gauge-invariant basis of Ref.~\cite{Drechsel:1997xv}:
\begin{equation}
\mathcal{M}^{\mu\nu}(p',q',p,q) =
\,e^2\sum\limits_{i=1}^{12}
\rho_i^{\mu\nu} A_i(\xi,t; q^2)\,,
\label{eq:vcs_amplitude}
\end{equation}
where $\mu\ (\nu)$ are the indices of the outgoing (incoming) photon four-vector fields,
with tensors $\rho_i$ given in Appendix~\ref{sec:appendix:tensors}.
The nucleon crossing symmetry in combination with charge conjugation yields the following property
of the invariant amplitudes,
\begin{subequations}
\begin{align}
A_i(\xi,t;q^2)&=+A_i(-\xi,t;q^2),\quad i=1,2,5,6,7,9,11,12\,,\\
A_i(\xi,t;q^2)&=-A_i(-\xi,t;q^2),\quad i=3,4,8,10\, ,
\label{eq:crossing}
\end{align}
\end{subequations}
with the latter equation leading to the above-mentioned reduction of the number of independent GPs
from 10 to 6. This property is also
helpful in checking our loop calculations.

The six GPs of Eq.~(\ref{eq:GPs}) are defined in terms of the non-Born amplitudes,\footnote{Expressions for the Born contribution in
this basis can for instance be found in Ref.~\cite{Pasquini:2001yy}.} 
\beq 
\bar A_i(\xi,t; q^2) \equiv  A_i(\xi,t; q^2) - A_i^\mathrm{Born}(\xi,t; q^2),
\eeq 
taken at $\xi=0$ and $t= q^2\equiv - Q^2$. Introducing a shorthand notation,
\beq 
\bar A_i(Q^2) \equiv \bar A_i(0,\,-Q^2; -Q^2), 
\eeq 
the precise expressions
for the GPs are given by
\begin{subequations}
\begin{align}
\label{eq:PL1L10}
P^{(L1,L1)0}(Q^2)&=\hphantom{-}\sqrt{\frac{2}{3}}N_q
\left[\bar{A}_1(Q^2)+4M^2(1+\tau)\bar{A}_2(Q^2)
+4M^2\tau\left(2\bar{A}_6(Q^2)+\bar{A}_9(Q^2)-\bar{A}_{12}(Q^2)\right)
\right]\,,\displaybreak[0]\\
\label{eq:PM1M10}
P^{(M1,M1)0}(Q^2)&=-\sqrt{\frac{8}{3}}N_q
\bar{A}_1(Q^2)\,,\displaybreak[0]\\
\label{eq:PL1L11}
P^{(L1,L1)1}(Q^2)&=-\frac{2}{3}N_q M\tau 
\left[\bar{A}_5(Q^2)+\bar{A}_7(Q^2)+4\bar{A}_{11}(Q^2)+4M\bar{A}_{12}(Q^2)
\right]\,,\displaybreak[0]\\
\label{eq:PM1M11}
P^{(M1,M1)1}(Q^2)&=-\frac{2}{3}N_q\frac{M\tau}{1+\tau}
\left[\bar{A}_5(Q^2)-2M\tau\bar{A}_{12}(Q^2)\right]\,,\displaybreak[0]\\
\label{eq:PM1L21}
P^{(M1,L2)1}(Q^2)&=\hphantom{-}\frac{2}{3}\sqrt{\frac{2}{3}}N_q
\left[\frac{\tau}{2(1+\tau)}\bar{A}_5(Q^2)+\frac{1}{2}\bar{A}_7(Q^2)+2\bar{A}_{11}(Q^2)
+\frac{M\tau}{1+\tau}\bar{A}_{12}(Q^2)
\right]\,,\displaybreak[0]\\
\label{eq:PL1M21}
P^{(L1,M2)1}(Q^2)&=-\frac{\sqrt{2}}{6}N_q
\frac{1}{1+\tau}
\left[8M \bar{A}_6(Q^2)+\bar{A}_7(Q^2)+4M\bar{A}_9(Q^2)+4\bar{A}_{11}(Q^2)+2M\tau\bar{A}_{12}(Q^2)
\right]
\,,
\end{align}
\end{subequations}
with the normalization factor $N_q$ usually taken to be
\beq 
N_q = \sqrt{\frac{2M+\nu_0}{2(M+\nu_0)}}= \sqrt{\frac{1+\tau}{1+2\tau}}\,.
\eeq

At the real-photon point, these GPs relate to the static nucleon
polarizabilities~\cite{Guichon:1995pu,Drechsel:1998zm}:
\begin{subequations}
\begin{align}
P^{(L1,L1)0}(0)&=-\mbox{$\frac{1}{\alpha_\mathrm{em}}\sqrt{\frac{2}{3}}$}\,\alpha_{E1}\,,
\label{eq:alphastatic}
\\
P^{(M1,M1)0}(0)&=-\mbox{$\frac{1}{\alpha_\mathrm{em}}\sqrt{\frac{8}{3}}$}\,\beta_{M1}\,,
\label{eq:betastatic}
\\
P^{(M1,L2)0}(0)&=-\mbox{$\frac{1}{\alpha_\mathrm{em}}
\frac{2}{3}\sqrt{\frac{2}{3}}$}\,\gamma_{M1E2}\,,
\label{eq:gM1E2static}
\\
P^{(L1,M2)0}(0)&=-\mbox{$\frac{1}{\alpha_\mathrm{em}}\frac{\sqrt{2}}{3}$}\,\gamma_{E1M2}\,,
\label{eq:gE1M2static}
\end{align}
\end{subequations}
where $\alpha_\mathrm{em}=e^2/4\pi\simeq 1/137$ is the fine structure constant.
The remaining two
GPs, $P^{(L1,L1)1}$ and $P^{(M1,M1)1}$, vanish (at $Q^2=0$). Their
slopes, on the other hand, can be related
to other static polarizabilities and to ``symmetric'' GPs  via the spin-dependent sum rules~\cite{Pascalutsa:2014zna,Hagelstein:2015egb,Lensky:2017dlc}.

The ``skewed'' generalizations  of the electric and magnetic dipole polarizabilities
are thus defined as follows:
\begin{subequations}
\begin{align}
\alpha_{E1}(Q^2)&=-\alpha_\mathrm{em}\mbox{$\sqrt{\frac{3}{2}}$}P^{(L1,L1)0}(Q^2)\,,\\
\beta_{M1}(Q^2)&=-\alpha_\mathrm{em}\mbox{$\sqrt{\frac{3}{8}}$} P^{(M1,M1)0}(Q^2)\,.
\end{align}
\end{subequations}
Similar generalizations can be made for the two spin 
polarizabilities in Eqs.\ (\ref{eq:gM1E2static}) and (\ref{eq:gE1M2static}). 

Finally, let us recall the relation to experimental observables.
As noted above, the six GPs of Eq.~(\ref{eq:GPs}) suffice to fully
parametrize the leading (linear in $\nu'$) term in the non-Born VCS amplitude.
The latter, together with the BH and Born VCS amplitudes, can be used
to calculate the observables of $ep$ bremsstrahlung. Most notably, the 
expression for the unpolarized five-fold differential cross-section can be cast in the following form,
\begin{equation}
\eqlab{LEX}
\mathrm{d}^5\sigma =\mathrm{d}^5\sigma^\mathrm{BH+Born}
+\nu'\Phi\left\{ V_1\left[P_{LL}(Q^2)-\frac{1}{\varepsilon}P_{TT}(Q^2)\right]
+V_2\sqrt{\varepsilon(1+\varepsilon)}P_{LT}(Q^2)
\right\}\,,
\end{equation}
where $\Phi$, $V_1$, and $V_2$ are kinematical factors (see Ref.~\cite{Guichon:1995pu} for the specific expressions thereof),
$\varepsilon$ is the electron polarization transfer parameter, $P_{LL}$,
$P_{TT}$, and $P_{LT}$ are the VCS response functions
given in terms of GPs as follows~\cite{Guichon:1998xv}:
\begin{subequations}
\eqlab{VCSresponse}
\begin{align}
P_{LL}(Q^2)&=-2\sqrt{6}M G_E(Q^2) P^{(L1,L1)0}(Q^2)\,,\label{eq:pll}\\
P_{TT}(Q^2)&=\hphantom{-}6 M G_M(Q^2)(1+\tau)
\left[
2\sqrt{2}\,M \tau\, P^{(L1,M2)1}(Q^2)+P^{(M1,M1)1}(Q^2)
\right]
\,,\label{eq:ptt}\\
P_{LT}(Q^2)&=\hphantom{-}\sqrt{\mbox{$\frac{3}{2}$}}M\sqrt{1+\tau}
\left[
G_E(Q^2)P^{(M1,M1)0}(Q^2)-\sqrt{6}\,G_M(Q^2)P^{(L1,L1)1}(Q^2)
\right]\,,\label{eq:plt}
\end{align}
\end{subequations}
where $G_E(Q^2)$ and $G_M(Q^2)$ are the Sachs electric and magnetic form factors of
the nucleon. 

The unpolarized differential cross-section thus gives information about two linear
combinations of the VCS response functions:
$\displaystyle P_{LL}-P_{TT}/\varepsilon$ and
$P_{LT}$. These two quantities are dominated by the scalar GPs: $P^{(L1,L1)0}$
and $P^{(M1,M1)0}$. Note, however, that performing unpolarized 
VCS experiments at fixed $Q^2$ and for two different values of $\varepsilon$
allows to separate $P_{LL}$ and $P_{TT}$ and thus to access one combination of
spin GPs, $P_{TT}$. To obtain information on all spin GPs, one
needs to consider polarization observables. We will be interested, in particular,
in the response function~\cite{Vanderhaeghen:1997bx}
\begin{align}
P^\perp_{LT}(Q^2)=\frac{M}{Q}\frac{G_E(Q^2)}{G_M(Q^2)}P_{TT}(Q^2)-\frac{Q}{4M}\frac{ G_M(Q^2)}{ G_E(Q^2)}P_{LL}(Q^2)\label{eq:pltperp}
\end{align}
which has been accessed experimentally using the beam-recoil polarization asymmetries \cite{Doria:2015dyx}.

\section{Chiral perturbation theory of generalized polarizabilities }\label{sec:BChPT}
Our aim is to compute the nucleon VCS amplitudes $A_i(\xi,t;q^2)$ and subsequently the ``skewed'' GPs
using the SU(2) chiral perturbation theory ($\chi$PT) \cite{Weinberg:1978kz,Gasser:1983yg}, 
including the nucleon and $\De(1232)$ 
degrees of freedom. We shall employ B$\chi$PT which is the manifestly-covariant extension of $\chi$PT to the single-baryon sector in its most straightforward
implementation (i.e., not the ``infrared regularization'' of Ref.~\cite{Becher:1999he}), where
the nucleon
is included as in\footnote{The power-counting concerns  
raised in \cite{Gasser:1987rb} have been overcome by renormalizing away the ``power-counting
violating'' using the low-energy constants (a.k.a., Wilson coefficients) available at that order.
 This has been shown explicitly within the  ``extended on-mass-shell renormalization scheme'' (EOMS)~\cite{Fuchs:2003qc}, 
 but is not limited to it.} Ref.~\cite{Gasser:1987rb}, and the $\De(1232)$ as in Ref.~\cite{Pascalutsa:2006up}, see also \cite{Geng:2013xn} for concise overview.
The heavy-baryon results can easily be obtained from B$\chi$PT by an additional expansion in the inverse powers of baryon masses.

\subsection{Further remarks on power counting}
Let us recall that the chiral effective-field theory is based on 
the perturbative expansion in powers of pion momentum $p$ and mass $m_\pi$ over the scale of spontaneous chiral symmetry breaking $\Lambda_\chi\sim 4\pi f_\pi$, with
$f_\pi\simeq 92$~MeV the pion decay constant. Each operator in the effective Lagrangian, or a
graph in the loopwise expansion of the S-matrix, can be assigned with an order of $p$.
To give a relevant example consider the operator 
\beq 
\bar N N F^2,
\eeq
with $N(x)$ standing for the Dirac field
of the nucleon, and $F^2$ is the square of the electromagnetic field tensor, $F_{\mu\nu}(x)=\pa_{[\mu}A_{\nu]}(x)$. This is an operator of $O(p^4)$, since two of the $p$'s come from the
photon momenta which are supposed to be small, and the other two powers arise because the two-photon
coupling to the nucleon must carry a factor of $\al_\mathrm{em}$ (the charge $e$ counts as $p$, as we want
the derivative of the pion field to count as $p$ even after including the minimal coupling to the photon).

This operator enters the effective Lagrangian with a so-called low-energy constant (LEC), denoted by $C$,
and it gives a contribution to the Compton scattering amplitude in the form of\footnote{
Throughout this paper we use the conventions summarized in the beginning of Ref.~\cite{Hagelstein:2015egb}.}
\beq 
\mathcal{M}^{\mu\nu}_C = C \, (q\cdot q' \, g^{\mu\nu} - q^\mu q^{\prime\,\nu}),
\eeq 
hence leading to a shift in the magnetic dipole polarizability: $\be_{M1}\to \be_{M1}+C/4\pi$.
Now, two important remarks are in order.
\begin{itemize}
\item[i)] {\em Naturalness}. The value of $C$ is not completely arbitrary, but should rather go as 
$C= (e^2/\La_\chi^3) c $, with the dimensional constant $c$ being of order of 1, or more precisely:
\beq 
p/\La_\chi \ll |c|\ll \La_\chi/p\,.
\eeq 
This condition ensures that the contribution  of this operator is indeed 
of $O(p^4)$, as the power counting commands.
\item[ii)] {\em Predictive powers}.
This LEC enters very prominently in Compton scattering and polarizabilities---at the tree level,
which means its value is best fixed by the empirical information on these quantities. 
If this is so, the $O(p^4)$ is not ``predictive'', as it could only be used to {\em fit} 
the $\chi$PT result to experiment or lattice QCD calculations. On the other hand, contributions
of orders lower than $p^4$ are predictive, as they only contain LECs fixed from elsewhere.
It is crucial to first study the predictive contributions, if there are any, and this is what we shall
focus on here, for the case of VCS and GPs.
\end{itemize}

The ``predictive'' contributions to Compton scattering and polarizabilities 
had been identified in Ref.~\cite{Lensky:2009uv} and computed  for the case
of real-Compton scattering therein and for VVCS in \cite{Lensky:2014dda}.
Our present calculation of VCS is quite analogous to those works and hence 
we refer to them for most of the technical details, such as the expressions for the
relevant terms of the effective Lagrangian.

On the conceptual side, it is important to note that the counting of the $\De(1232)$ effects is done in the so-called ``$\delta$-counting''~\cite{Pascalutsa:2002pi}. In it, the Delta-nucleon mass difference
$\varDelta=M_\Delta-M$ is counted as a light scale ($\vDe \ll \La_\chi$) which is
substantially heavier than the pion mass ($m_\pi\ll\vDe)$. Hence, if $p\sim m_\pi$, 
then $O(p^4/\vDe)$ is in between of $O(p^3)$ and $O(p^4)$.

\begin{figure}[ht]
 \includegraphics[width=0.6\columnwidth]{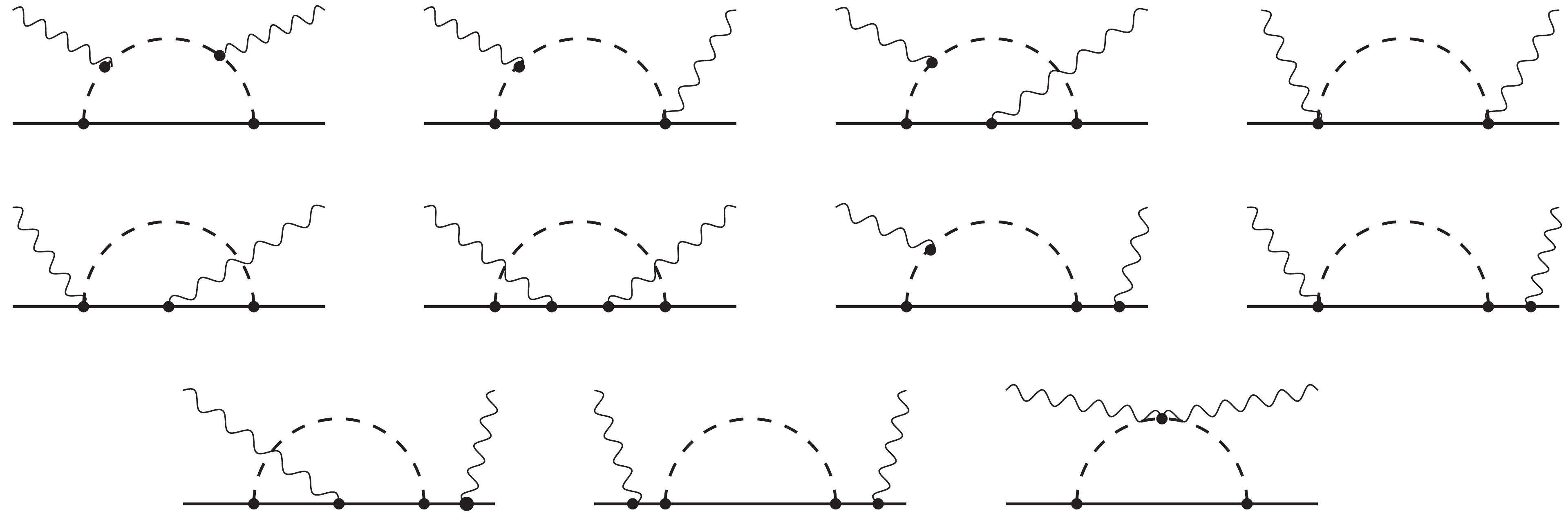}
\caption{Pion-nucleon loops of $O(p^3)$.
Solid (dashed) lines denote nucleons (pions). 
Crossed and time-reversed graphs are not shown but are included in the calculation.}
\label{fig:piNloops}
\end{figure}

\begin{figure}[hbt]
\includegraphics[width=0.6\columnwidth]{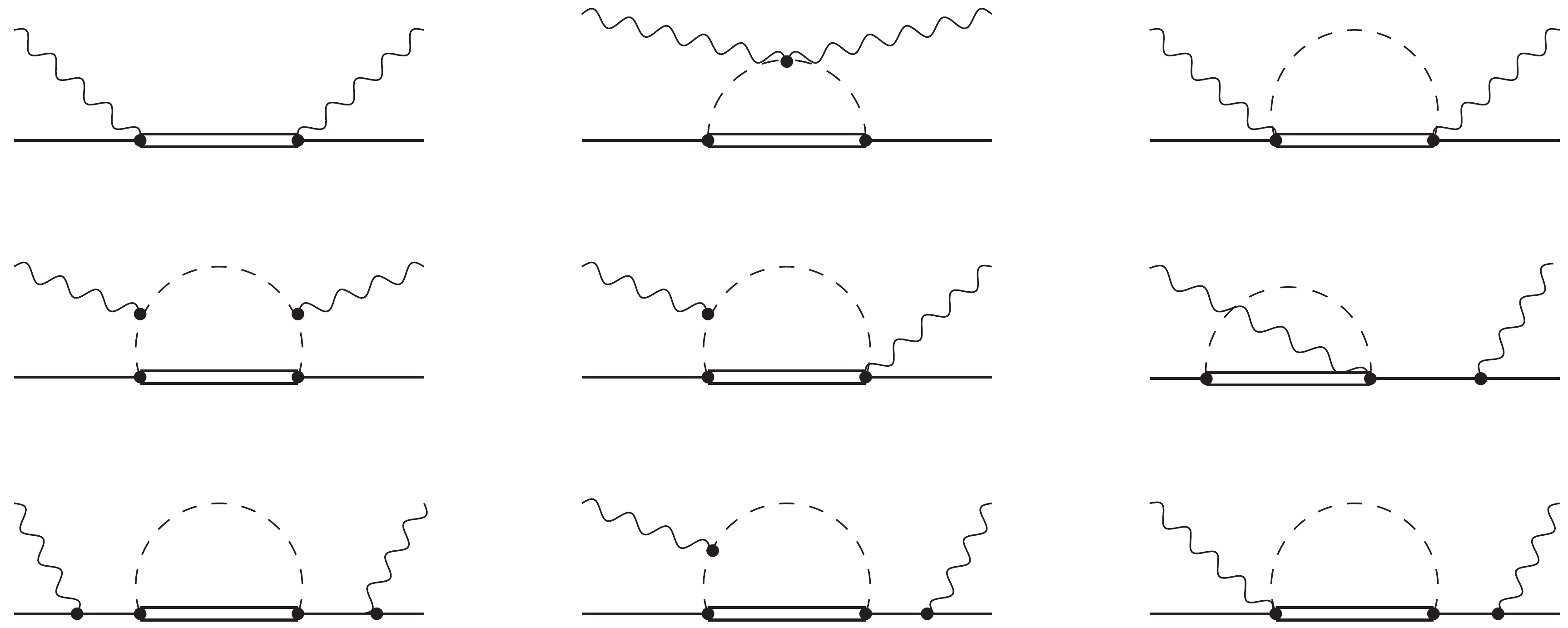}
\caption{Pion-Delta loops and the Delta tree of $O(p^4/\varDelta)$ (in the low-energy regime).
Crossed and time-reversed graphs are not shown but are included in the calculation. 
Double lines denote the propagator of the Delta.}
\label{fig:piDloops}
\end{figure}

For the non-Born VCS amplitude
and polarizabilities the predictive orders are $O(p^3)$ and $O(p^4/\vDe)$.
The $O(p^3)$ contribution comes from the pion-nucleon loops shown in Fig.~\ref{fig:piNloops}.
We refer to it here as the leading-order (LO) contribution.\footnote{In the full Compton amplitude (i.e., 
including the Born term), it is, in fact, a next-to-leading order contribution, and this is how it is referred to sometimes, e.g.~\cite{Lensky:2009uv}. }
The $O(p^4/\vDe)$ contribution, arising at the next-to-leading order (NLO), 
comes from the Delta pole graph and the pion-Delta loops shown in Fig.~\ref{fig:piDloops}.


Going into further detail, we note that 
the feature of the $\delta$-counting is that the characteristic momentum
$p$ distinguishes two regimes: {\it low-energy} ($p\simeq m_\pi$), and 
{\it resonance} ($p\simeq \varDelta$). The above counting is limited to the 
low-energy regime.
Since we are interested in the VCS
amplitude at the specific kinematics point where the GPs are
defined (i.e., $\xi=0$, $t=-Q^2$), we do not consider the regime
where one-Delta-reducible graphs are enhanced (resonance regime).
However, going to higher $Q$ one does need to count the 
Delta propagators similar to the nucleon propagators, which, in turn, calls for inclusion
of pion-Delta loops with two and three Delta propagators, which have been omitted here. They are only included implicitly to restore current conservation
 by adjusting the isospin coefficients of one-nucleon-reducible graphs in Fig.~\ref{fig:piDloops}, as explained in Ref.~\cite{Lensky:2009uv}. Apart from that, pion-Delta loops have rather mild
dependence on momenta and the missing loops are unlikely to significantly affect the
$Q^2$-dependence of the GPs, even for $Q^2$ comparable to $\vDe^2$. 

To conclude this section, a remark is in order about the $\pi^0$ anomaly graph, sometimes considered
to be a part of the Born contribution.
It enters the VCS amplitude at $O(p^2)$ and represents the dominant part of
the spin GPs (all except $P^{(L1,L1)1}$ where it does not enter).
However, the anomaly contributions cancel in the response functions
introduced above. We will also omit them when showing results for the spin GPs.

\subsection{In practice}

The calculation of the $\pi N$ and $\pi\Delta$ loop graphs in Fig.~\ref{fig:piNloops} and \ref{fig:piDloops} is analogous to Ref.~\cite{Lensky:2009uv}, with the obvious extension to the case of finite virtuality of the initial photon. The renormalization is done in the exactly the same way; 
namely, graphs with the nucleon self-energy
and with the one-loop $\gamma NN$ vertices are subtracted according to
the usual prescription
\begin{align}
\Sigma_R(\slashed{p}_s)&=\Sigma(\slashed{p}_s)-\Sigma(\slashed{p})-\Sigma'(\slashed{p})(\slashed{p}_s-M)\,,\\
\Gamma^\mu_R(p,p')&=\Gamma^\mu(p,p')-\Big[\gamma^\mu F_1(q^2)
-\frac{1}{2M}\gamma^{\mu\nu}q_\nu F_2(q^2)
\Big]\,,
\end{align}
where $p_s$ and $p$ in the first equation are the off-shell and the on-shell momentum of the nucleon, whereas
$F_1(q^2)$ and $F_2(q^2)$ are the on-shell nucleon Dirac and Pauli form factors
resulting from the unsubtracted $\gamma NN$ vertex $\Gamma^\mu(p,p')$, with $q=p'-p$
being the momentum transfer from the photon to the nucleon.

The Delta pole graph in Fig.~\ref{fig:piDloops} is calculated 
in Refs.~\cite{Lensky:2014dda,Pascalutsa:2005vq}, and as in those works
the magnetic $\gamma N\Delta$ coupling $g_M$ acquires the dipole behavior that mimics the form expected from 
vector-meson dominance:
\begin{equation}
g_M\to \frac{g_M}{\big[1+\left(Q/\Lambda\right)^2\big]^2}\,,
\end{equation}
with the dipole mass $\Lambda^2=0.71$~GeV${^2}$.

Concerning the implementation of  tensor decomposition in Eq.~(\ref{eq:vcs_amplitude}), 
it proved to be useful to write the basis tensors $\rho_i$,
$i=1,\dots,12$, in terms of the Tarrach tensors $T_i$, $i=1,\dots,34$,
introduced for the most general VVCS case~\cite{Tarrach:1975tu}, see
Appendix~\ref{sec:appendix:tensors}.
All tensors $\rho_i$, apart
from $\rho_2$, $\rho_3$, and $\rho_6$, have unique structures that allow for unambiguous identification of the corresponding parts of the amplitude,
e.g., the combination $T_{29}+T_{30}$ enters only $\rho_{12}$, $T_{17}$ enters only $\rho_{11}$, and so on.
After the tensors $\rho_i$, $i\neq 2,3,6$, have been identified,
the remaining tensors can be identified as well, since at this stage they
are the only ones that can enter the rest of the amplitude.
Since the basis $\rho_i$ is explicitly gauge invariant, all the terms that
are not proportional to any of $\rho_i$ have to vanish when one decomposes
a gauge invariant amplitude, e.g., summing up a gauge invariant subset of Feynman graphs such as the $\pi N$ loops in Fig.~\ref{fig:piNloops} with their crossed and time-reversed partners, or the Delta pole graph in Fig.~\ref{fig:piDloops}
with its crossed partner, or the $\pi \Delta$ loops in that figure with their
crossed and time-reversed partners. Ensuring that the rest of the amplitude
vanishes after the terms proportional to $\rho_i$ have been subtracted
represents a non-trivial check of a VCS calculation. 

We note as well that the tensor
decomposition introduces false singularities in the amplitudes due to
some of the coefficients in front of $T_i$ proportional to $\xi$; these
singularities disappear in the end, which serves as yet another check of the calculation.
These singularities tend to interfere with the false on-shell singularities
in one-nucleon-reducible graphs, i.e., graphs with the nucleon self-energy
loop and those with the one-loop $\gamma NN$ vertices. These latter singularities
also have to disappear in the end since both the self-energy and the one-loop
$\gamma NN$ vertices are subtracted on-shell, as explained above. 

It
is more convenient, however, to explicitly remove on-shell singularities from
the integrals over the Feynman parameters; this can be done by integrating by
parts in these integrals and by additional subtractions where they are needed.
To illustrate these techniques, we give two examples of typical terms
arising in the $\pi N$ self-energy graph,
where the on-shell singularities are manifest:
\begin{align}
\text{A)}\ X_1&=\frac{1}{\bar{s}-1}\int\limits_{0}^{1}\mathrm{d}x
\left\{
\log\left[x^2+\mu^2(1-x)-(\bar{s}-1)x(1-x)\right]-\log\left[x^2+\mu^2(1-x)\right]\right\}\,,\\
\text{B)}\ X_2&=
\frac{1}{\bar{s}-1}\int\limits_{0}^{1}\mathrm{d}x\frac{x^2 \left[x^2 (2x-1)-3 \mu ^2 (x-1)^2+2(\bar{s}-1)x (x-1)^2\right]}
{
   \left[x^2+\mu^2(1-x)\right]
   \left[x^2+\mu^2(1-x)-(\bar{s}-1)x(1-x)\right]}
\,,
\end{align}
here $\mu=m_\pi/M$ and $\bar{s}=s/M^2$. The singularity at $\bar{s}\to 1$
appearing in $X_1$ is cancelled when one integrates by parts in the first integral,
whereas in order to deal with $X_2$ one notices that
the integral in it vanishes at $\bar{s}\to 1$. This means that the integrand
of $X_2$ can be subtracted at $\bar{s}=1$; the singularity cancels after this subtraction.

Removing the on-shell false singularities explicitly
allows one to deal with the remaining $1/\xi$ false singularities that come
from the tensor decomposition by simply expanding the integrals in powers of $\xi$.
It appears to be possible to analytically verify that coefficients
in front of negative powers of $\xi$ turn to zero after integration
over the Feynman parameters, both for the $\pi N$ and $\pi \Delta$ loops.

Given the kinematics of VCS, one is only interested in
the $\xi^0$ term in the expansion of the thus obtained amplitudes
$A_i(\xi, t; q^2)$, while the Mandelstam variable $t$ is also
set to $t=q^2\equiv-Q^2$. The resulting functions $\bar{A}_i(Q^2)$ are obtained
as integrals over two Feynman parameters; an expansion in $Q^2$ around
the static point $Q^2=0$ allows for the integrals to be taken analytically.
Appendix~\ref{sec:appendix:amplitudes} contains expressions for those
linear combinations of $\bar{A}_i(Q^2)$ that enter the GPs given by Eq.~\ref{eq:GPs},
resulting from the $\pi N$ loop graphs in Fig.~\ref{fig:piNloops} and from the
Delta pole graph in Fig.~\ref{fig:piDloops}. The corresponding expressions for
the $\pi \Delta$ loops in Fig.~\ref{fig:piDloops} are given in the supplementary material to this article.

\subsection{Error estimate}\label{sec:error}
In making comparison with experimental data, it is important to provide
a theoretical uncertainty. In the case of an EFT expansion,
the common way to obtain this uncertainty is via the estimate of
higher-order contributions. This work employs the following estimate.
In the low-momenta regime, where the expansion
parameter is $\displaystyle\delta\sim p/\varDelta$, our calculation is of
the next-to-leading order (NLO). A conservative estimate of the next-to-next-to-leading order (NNLO) contributions would be $\text{error}(f)=\delta^2 f$,
where $f$ is a generic VCS amplitude or response function.
It is important to note, however, that the error of the scalar polarizabilities
$\alpha_{E1}(Q^2)$ and $\beta_{M1}(Q^2)$ in the static limit $Q^2=0$ is
defined by the error of the corresponding static (real) polarizabilities.
This error was argued to be small~\cite{Lensky:2015awa}
due to the fact that these polarizabilities
are very close at NLO to the results obtained in B$\chi$PT fits to
real Compton scattering data~\cite{Lensky:2014efa}, and that there are contact terms at NNLO that
will in any case compensate changes in $\alpha_{E1}$ and $\beta_{M1}$ coming
from other higher-order mechanisms. The static errors are estimated as
$\text{error}(\alpha_{E1},\text{static})\simeq\text{error}(\beta_{M1},\text{static})\sim 0.7\times 10^{-4}$~fm${^3}$
(see Ref.~\cite{Lensky:2015awa}); this
translates to the uncertainty of $4.7$~GeV${^2}$ and $2.3$~GeV${^2}$ in $P_{LL}(0)$ and
$P_{LT}(0)$, respectively. This static uncertainty has to dominate at very small
$Q^2$, whereas at larger $Q^2$ (still in the low-momenta regime) the
term $\delta^2 f\sim (p^2/\varDelta^2) f$ will become more important.
In practice, we take the sum of the two values.

The uncertainty estimate in the high-momenta regime works in a similar way.
In this regime, $p\gtrsim\varDelta$, our calculation is at an incomplete leading 
order (LO), however, we will treat it as an LO calculation as argued above.
The expansion parameter in this regime can be one of these,
\beq\nonumber
\delta
=\left\{\frac{m_\pi}{\varDelta},\frac{p}{\Lambda_\chi},\frac{\varDelta}{\Lambda_\chi}\right\}\,,
\eeq
and we take the average value of the three in order to
estimate the NLO contribution. In summary, our uncertainty estimate for
a VCS amplitude or a response function $f$ is given by
\begin{equation}
\text{error}(f)=
\left\{
\begin{matrix}
\text{error}(f,\text{static})+\displaystyle\frac{p^2}{\varDelta^2}f\,, & p\lesssim \varDelta\,,  \\
& \\
\displaystyle\frac{1}{3}\left(\frac{m_\pi}{\varDelta}+\frac{p}{\Lambda_\chi}+\frac{\varDelta}{\Lambda_\chi}\right)f\,, & p\gtrsim\varDelta\,,
\end{matrix}
\right.
\end{equation}
where $\text{error}(f,\text{static})$
is the static uncertainty discussed above. To obtain
smooth bands in the plots, the uncertainties in the different regimes
are multiplied by smooth transition functions. One has to note that
this error estimate can lead to artifacts such as zero crossings,
in the regime $p\gtrsim\varDelta$: the error
being proportional to the observable, it can become small or
even turn to zero if the latter decreases or vanishes.
While we still consider this issue to be tolerable as far as the plots we demonstrate
here are concerned, one can see it manifest,
for instance, in Figs.~\ref{fig:response_function_PLT} and~\ref{fig:alpha_beta}
below, where the bands of $P_{LT}$ and $\beta_{M1}$ shrink at larger values of $Q^2$.

\section{Comparison with previous calculations}\label{sec:comparison}

In this section, we compare our B$\chi$PT results with previous results
obtained in HB$\chi$PT, in the linear sigma model, and with fixed-$t$
dispersion relations.
Matching our results against those obtained in the former two frameworks
provides an important check of our calculation.

\subsection{Linear $\si$-model}\label{sec:comparision:lsm}
The first check is made by comparing our results
with the results of Metz and Drechsel~\cite{Metz:1996fn,Metz:1997fr}
who calculated the nucleon GPs in the linear sigma model at one-loop level.
Their linear sigma model calculation, performed in the limit of infinitely large sigma
meson mass, is exactly equivalent to the $O(p^3)$
one-loop pion-nucleon B$\chi$PT result. The easiest way to see it is perhaps to
compare the Lagrangian used in Refs.~\cite{Metz:1996fn,Metz:1997fr} with the
B$\chi$PT Lagrangian after the field redefinition done in Ref.~\cite{Lensky:2009uv}.
Metz and Drechsel provide only the expressions
at $Q^2=0$ for the GPs and their derivatives. They also give second derivatives for
the two spin-dependent GPs that vanish at $Q^2=0$. The expressions
for all of the spin-dependent GPs are in addition expanded in $1/M$ up to NNLO. 
Our calculation has been able to reproduces all of their results except one:
their expression for $\alpha_{E1}$ of the proton, which we believe to be
due to a typo in the second line of Eq.~(17) of Ref.~\cite{Metz:1996fn},
namely, the first term in the square brackets should read $152$ instead of $157$.

\subsection{Heavy-baryon expansion}\label{sec:comparison:hb}
By expanding our results in powers of $1/M$ we can check against 
the HB$\chi$PT calculation of Hemmert et al.~\cite{Hemmert:1999pz}
that includes the Delta
isobar in the $\eps$-expansion~\cite{Hemmert:1996xg}. We checked that
the leading term of the heavy-baryon expansion of our results for the scalar GPs
(with the $\pi N$ and $\pi\Delta$ loops corresponding to, respectively, $O(p^3)$
and $O(\epsilon^3)$ result of Hemmert et al.)
reproduces the results of Ref.~\cite{Hemmert:1999pz}.

The spin-dependent GPs have also been calculated in HB$\chi$PT without
the Delta isobar up to incomplete $O(p^5)$ in Refs.~\cite{Kao:2002cn,Kao:2004us}.
These calculations include $\pi N$ loops with photon couplings to the anomalous magnetic
moment (a.m.m.) of the nucleon inside the loop, which appear at $O(p^4)$ and
are not included
in our calculation. Nevertheless, the heavy-baryon expansion of our results
should reproduce their HB$\chi$PT expressions, once the a.m.m.\ couplings are
set to zero. We have reproduced the HB$\chi$PT
expressions for $P^{(M1,M1)1}$ and $P^{(L1,M2)1}$, calculated
in Ref.~\cite{Kao:2004us}. The other two spin-dependent GPs, $P^{(L1,L1)1}$ and $P^{(M1,L2)1}$, 
are reproducible up to the leading order in $1/M$; the differences at NLO
start in the second non-vanishing terms in the expansion in powers of $Q^2$,
i.e., the first derivative of $P^{(M1,L2)1}$ with respect to $Q^2$,
and in the second derivative of $P^{(L1,L1)1}$. 

Is is important to realize that $P^{(L1,L1)1}$ and $P^{(M1,L2)1}$ at $O(p^5)$ were deduced
in Ref.~\cite{Kao:2004us}  by using the nucleon 
crossing relations of \Eqref{crossing}, which in HB$\chi$PT do not hold
exactly due to the lack of an exact charge conjugation symmetry.  The mismatch between the HB results
and ours demonstrates that these crossing relations should not be used in HB$\chi$PT 
to obtain complete expressions for the terms of higher-order in $1/M$ expansion.

\begin{figure}[th]
\includegraphics[width=\textwidth]{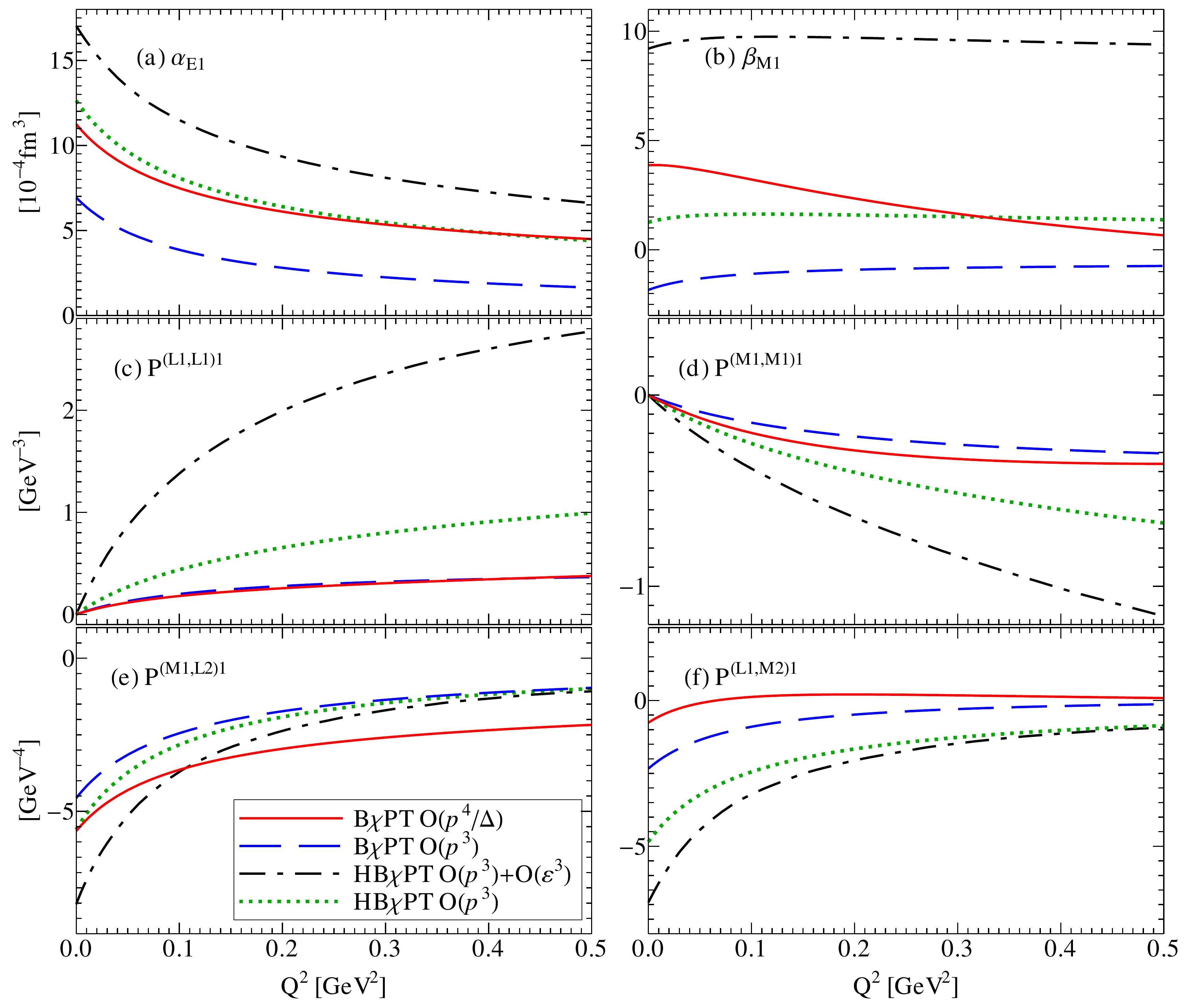}
\caption{Comparison of covariant B$\chi$PT and HB$\chi$PT
results for generalized polarizabilities:
(a) $\alpha_{E1}(Q^2)$; (b) $\beta_{M1}(Q^2)$; (c) $P^{(L1,L1)1}(Q^2)$;
(d) $P^{(M1,M1)1}(Q^2)$; (e) $P^{(M1,L2)1}(Q^2)$; (f) $P^{(L1,M2)1}(Q^2)$.
Red solid curve: covariant $O(p^3)+O(p^4/\varDelta)$;
blue dashed curve: covariant $O(p^3)$;
black dot-dashed curve: HB $O(p^3)+O(\epsilon^3)$;
green dotted curve: HB $O(p^3)$. HB results are from Refs.~\cite{Hemmert:1999pz,Kao:2004us}.
}
\label{fig:gps}
\end{figure}
For completeness, we provide here our results for the HB expansion
of the spin GPs up to NLO in $1/M$. The LO results are the same
as given in Refs.~\cite{Hemmert:1999pz,Kao:2002cn,Kao:2004us} and read
\begin{subequations}
\begin{align}
P^{(L1,L1)1}_\mathrm{HB,LO}(Q^2) &=\hphantom{-}
\frac{g_A^2 Q^2}{288 \pi ^2 f_\pi^2 m_\pi^2 M}
 \left[3- \left(2 w^2+1\right)L_1(w)\right]\,,\displaybreak[0]\\
P^{(M1,M1)1}_\mathrm{HB,LO}(Q^2) &=-\frac{g_A^2 Q^2}{288 \pi ^2 f_\pi^2 m_\pi^2 M}
\left[3-2 \left(w^2+1\right)L_1(w)\right]\,,\displaybreak[0]\\
P^{(M1,L2)1}_\mathrm{HB,LO}(Q^2) &=-\frac{g_A^2}{72 \sqrt{6} \pi ^2 f_\pi^2 m_\pi^2} L_1(w)\,,\displaybreak[0]\\
P^{(L1,M2)1}_\mathrm{HB,LO}(Q^2) &=-\frac{g_A^2}{144 \sqrt{2} \pi^2 f_\pi^2 m_\pi^2} L_1(w)\,,
\end{align}
\end{subequations}
with nucleon axial coupling constant $g_A \simeq 1.27$, pion decay constant $f_\pi \simeq 92.21$~MeV, 
where  $w=Q/2m_\pi$,
and the function $L_1(w)$ is defined as
\beq 
L_1 (w)=\frac{3}{2w^2}\left(1-\frac{\sinh ^{-1}w}{w \sqrt{w^2+1}}\right), \quad L_1(0)=1\,.
\eeq 
The correct NLO results (without the nucleon a.m.m.\ couplings) read
\begin{subequations}
\begin{align}\scriptsize
P^{(L1,L1)1}_\mathrm{HB,NLO}(Q^2) &=-\frac{g_A^2 Q^2 }{2304 \pi f_\pi^2 m_\pi M^2 }\left[
\frac{27 w^2+30}{ w^2+1}
-(9 w^2+4)L_2(w)
+ \tau_3 \left(9-(3 w^2+1)L_2(w)\right)\right]
\displaybreak[0]
\,,\\
P^{(M1,M1)1}_\mathrm{HB,NLO}(Q^2) &=\hphantom{-}\frac{g_A^2 Q^2 }{768 \pi  f_\pi^2 m_\pi M^2}\left[5 -
\frac{5 w^2+3}{3} L_2(w)+ \tau_3 \left(1 - \frac{1}{3}(w^2+1)L_2(w)\right)\right]
\displaybreak[0]
\,,\\
P^{(M1,L2)1}_\mathrm{HB,NLO}(Q^2) &=\hphantom{-}\frac{g_A^2}{576 \sqrt{6} \pi  f_\pi^2 M m_\pi}\left[-
\frac{3 w^2}{w^2+1}+
(w^2+2) L_2(w)
+ \tau_3 \left(3+(1-w^2)L_2(w)\right)\right]
\displaybreak[0]
\,,\\
P^{(L1,M2)1}_\mathrm{HB,NLO}(Q^2) &=\hphantom{-}\frac{g_A^2}{768 \sqrt{2} \pi  f_\pi^2 M m_\pi}
\left[\frac{5 w^2+ 7}{w^2+1}
-\frac{5w^2-3}{3} L_2(w)
+  \tau_3 \left(1 + \frac{1}{3}(1-w^2)L_2(w)\right)\right]\,,
\end{align}
\end{subequations}
where
\beq
L_2(w)=\frac{3 \left(w-\tan ^{-1}w\right)}{w^3},\quad L_2(0)=1\,,
\eeq
and $\tau_3=+1$ or $-1$ for the proton or the neutron, respectively.

Our B$\chi$PT results for the GPs are furthermore compared with the analogous HB$\chi$PT
results in Fig.~\ref{fig:gps}. Panels (a) and (b) show, respectively, the results
for $\alpha_{E1}(Q^2)$ and $\beta_{M1}(Q^2)$; one can see that,
while the HB$\chi$PT $O(p^3)$ $\pi N$ loops give results
very similar to the full B$\chi$PT result (which describes the data quite
well, as discussed above), the Delta isobar contribution
at $O(\epsilon^3)$ is simply too large to provide a reasonable description
of the data. On the other hand, the B$\chi$PT $O(p^3)$ $\pi N$ loops
underpredict the scalar GPs, which helps to accommodate the Delta isobar contribution
at $O(p^4/\varDelta)$.

A similar pattern emerges in the case of the spin-dependent GPs, shown
in panels (c)-(f); the two GPs that vanish at $Q^2=0$, $P^{(L1,L1)1}$ and $P^{(M1,M1)1}$, are much larger in HB$\chi$PT, especially with the Delta isobar.
The differences between B$\chi$PT and HB$\chi$PT are perhaps not that
large for one of the remaining two GPs, $P^{(M1,L2)1}$,
whereas $P^{(L1,M2)1}$ differs more significantly. This can be
traced to the values of the spin polarizability $\gamma_{E1M2}$
being different in B$\chi$PT and HB$\chi$PT; one has to note, however, that
this pattern will change once higher orders in the expansion are included
(see also the discussion below).

\subsection{Fixed-$t$ dispersion relations}\label{sec:comparison:dr}
\begin{figure}[th]
\includegraphics[width=\textwidth]{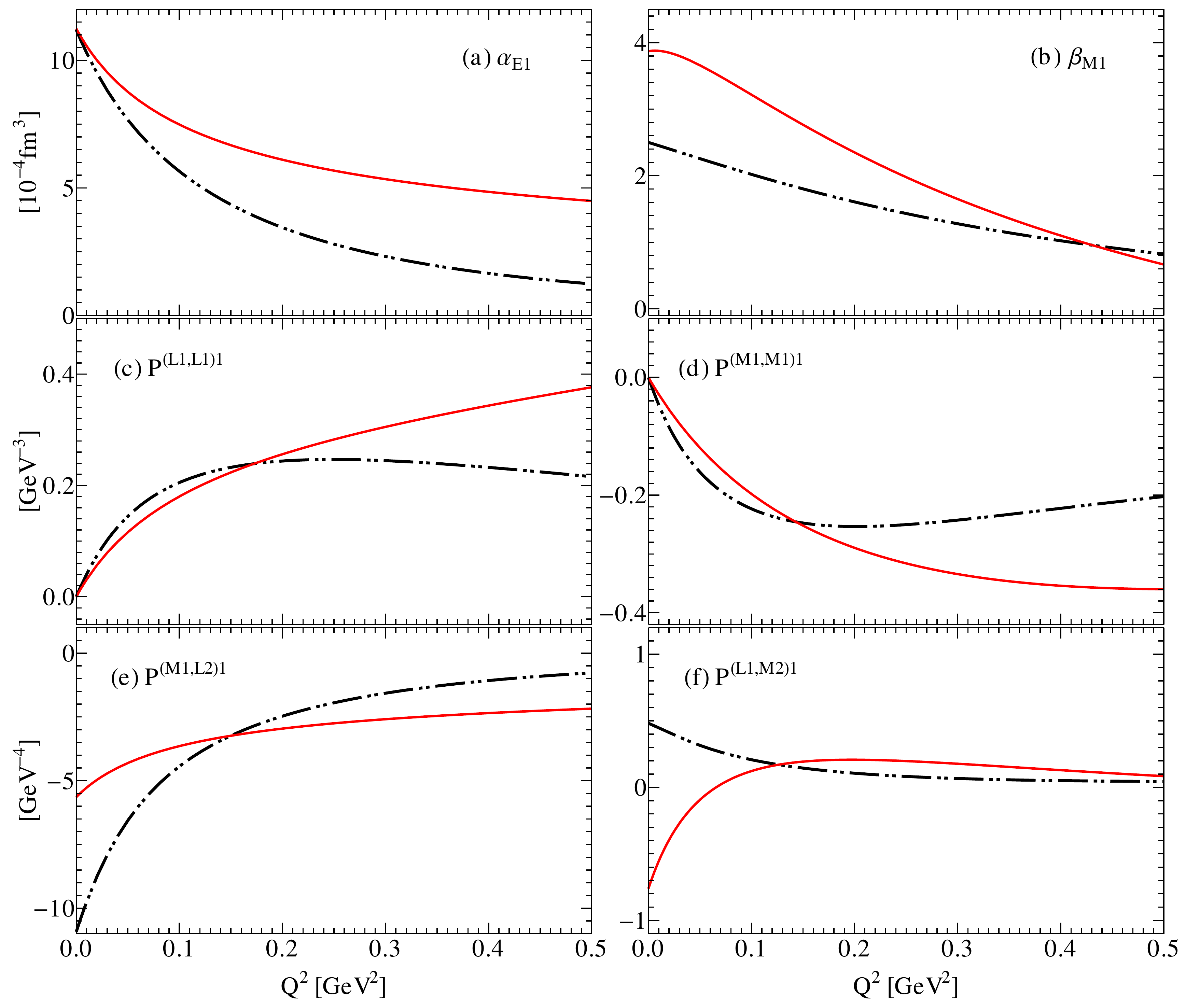}
\caption{Comparison of covariant B$\chi$PT
and DR results for generalized polarizabilities:
(a) $\alpha_{E1}(Q^2)$; (b) $\beta_{M1}(Q^2)$; (c) $P^{(L1,L1)1}(Q^2)$;
(d) $P^{(M1,M1)1}(Q^2)$; (e) $P^{(M1,L2)1}(Q^2)$; (f) $P^{(L1,M2)1}(Q^2)$.
Red solid curve: covariant B$\chi$PT $O(p^3)+O(p^4/\varDelta)$;
black dash-double-dotted curve: DR~\cite{Drechsel:2002ar}.
}
\label{fig:gpsDR}
\end{figure}

We finally compare our results with the calculations based on 
fixed-$t$ dispersion relations (DR) for the VCS amplitudes~\cite{Drechsel:2002ar}.
In Fig.~\ref{fig:gpsDR} we compare the numerical results
for the proton GPs. The fixed-$t$ DR calculations rely on the empirical input 
of pion electro-production multipoles. We compare here with the updated results
of Ref.~\cite{Drechsel:2002ar} based on the MAID-2007~\cite{Drechsel:2007if} pion electroproduction multipole analysis.

Panel (a) of Fig.~\ref{fig:gpsDR} shows the electric polarizability, for which 
one can see a very good agreement at $Q^2=0$, which quickly worsens with increasing $Q^2$. 
For the magnetic polarizability, one sees quite an opposite picture, see
panel (b). The current PDG value for the static magnetic polarizability,  
$\beta_{M1}=2.5(4)\times 10^{-4}$~fm${^3}$, is adopted in the fixed-$t$ DR
result. Our B$\chi$PT prediction is substantially larger~\cite{Lensky:2009uv,Lensky:2015awa}: $\beta_{M1}=3.9(7)$ in the usual units. 
Fits of Compton scattering data based on $\chi$PT also tend to yield a larger value~\cite{Griesshammer:2012we,Lensky:2014efa}: $\beta_{M1}\simeq 3.2(5)$.

As for higher $Q^2$, the $t$DR calculation of Ref.~\cite{Drechsel:2002ar} imposes a dipole fall-off
of the subtraction function in the scalar polarizabilities: 
\begin{align}
\label{eq:dipole_alpha}
\alpha_{E1}^\mathrm{DR}(Q^2)&=\alpha_{E1}^{\pi N}(Q^2)+\frac{\alpha_{E1}-\alpha_{E1}^{\pi N}}{\bigl[1+Q^2/\Lambda_\alpha^2\bigr]^2}\,,\\
\label{eq:dipole_beta}
\beta_{M1}^\mathrm{DR}(Q^2)&=\beta_{M1}^{\pi N}(Q^2)+\frac{\beta_{M1}-\beta_{M1}^{\pi N}}{\bigl[1+Q^2/\Lambda_\beta^2\bigr]^2}\,,
\end{align}
where $\alpha^\mathrm{DR}_{E1}(Q^2)$ is the full DR result,
$\alpha^{\pi N}_{E1}(Q^2)$ is the $\pi N$ contribution, with $\alpha_{E1}$ and
$\alpha_{E1}^{\pi N}$ being the corresponding values at $Q^2=0$, with the
analogous definitions for $\beta_{M1}(Q^2)$.
In using the $t$DR results we fix the static values of $\{\alpha_{E1},\beta_{M1}\}$
to the current PDG values $\{11.2,2.5\}\times 10^{-4}$~fm${^3}$, whereas
the cut-offs
$\{\Lambda_\alpha,\Lambda_\beta\}=\{0.631\pm 0.011,0.745\pm 0.021\}$~GeV
are taken from the recent fit of VCS data~\cite{Correa:thesis}.

For the spin polarizabilities,
the GPs $P^{(L1,L1)1}$
and $P^{(M1,M1)1}$ (panels (c) and (d) of Fig.~\ref{fig:gpsDR}, respectively), which vanish for real photons, 
show a good agreement between B$\chi$PT and DR,
especially at low $Q^2$. The agreement for $P^{(M1,L2)1}$ and $P^{(L1,M2)1}$,
shown in panels (e) and (f) of that figure,
is not so good. Especially for $P^{(L1,M2)1}$ one notices a different slope at $Q^2=0$
between the B$\chi$PT and DR results. On the other hand, $P^{(M1,L2)1}$ and $P^{(L1,M2)1}$
correspond, in the limit $Q^2=0$, to the two mixed spin polarizabilities
$\gamma_{M1E2}$ and $\gamma_{E1M2}$ (see Eqs.~\ref{eq:gM1E2static}-\ref{eq:gE1M2static}). The former is about two times larger in
DR than in B$\chi$PT~\cite{Lensky:2015awa}, which would explain the differences in $P^{(M1,L2)1}$ at low $Q^2$. The second is small and not well
constrained, which means that the difference between DR and B$\chi$PT
is probably not a very serious issue at this stage.
\begin{table}[htb]
\begin{tabular}{||c||c|c||}\hline
  Source                                           & $\gamma_{M1E2}$& $\gamma_{E1M2}$   \\
\hline
B$\chi$PT~\cite{Lensky:2015awa}                    &  $1.1 \pm0.3 $ &  $ 0.2\pm 0.2 $      \\
\hline
Fixed-$t$ DR~\cite{Drechsel:2002ar,Pasquini:2007hf}
                                                   &  $2.2       $  &  $-0.1$                   \\
\hline
HB$\chi$PT~\cite{McGovern:2012ew,Griesshammer:2015ahu}
                                                   &  $1.9\pm0.5$   & $-0.4\pm0.6$                         \\
\hline
MAMI 2015~\cite{Martel:2014pba}
                                                   & $1.99\pm 0.29$ & $-0.7\pm 1.2$           \\
\hline
\end{tabular}
\caption{Values of proton mixed spin polarisabilities $\gamma_{E1M2}$ and $\gamma_{M1E2}$, in units of $10^{-4}$~fm$^{4}$
resulting in the different frameworks: $O(p^4/\varDelta)$ B$\chi$PT~\cite{Lensky:2015awa}, fixed-$t$ DR~\cite{Drechsel:2002ar,Pasquini:2007hf} based on the MAID-2007~\cite{Drechsel:2007if} multipoles,
and $O(p^4)$ HB$\chi$PT~\cite{McGovern:2012ew,Griesshammer:2015ahu}, compared with 
the latest empirical extraction from experimental data~\cite{Martel:2014pba}.
}
\label{tab:polsspin}
\end{table}
 To further illustrate this point, we show in Table~\ref{tab:polsspin} the values of the two mixed
polarizabilities, $\gamma_{M1E2}$ and $\gamma_{E1M2}$, resulting in B$\chi$PT framework at $O(p^4/\varDelta)$, in fixed-$t$ DR,
in HB$\chi$PT at $O(p^4)$, and the results of extraction
of the spin polarizabilities from experimental data of one of the 
beam-target asymmetries, $\Sigma_{2x}$.
 
\section{Results for VCS observables}\label{sec:results}

\begin{figure}[b]
\includegraphics[width=0.6\textwidth]{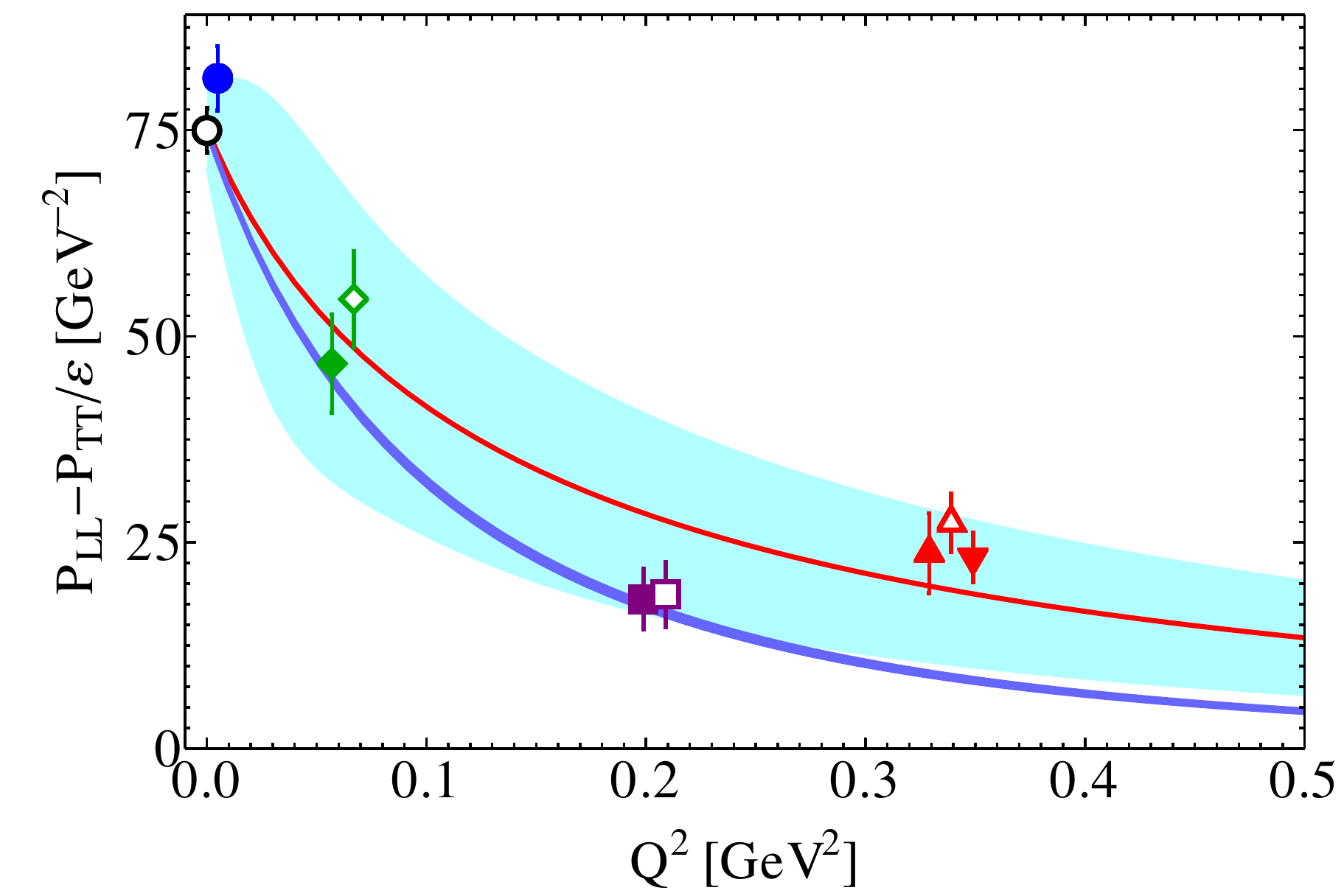}
\caption{VCS response function $\displaystyle P_{LL}(Q^2)-P_{TT}(Q^2)/\varepsilon$.
The total $O(p^3)+O(p^4/\varDelta)$ result is given by the red solid curve
with the cyan band showing the
estimated theoretical uncertainty as explained in the text. DR results~\cite{Drechsel:2002ar}
are shown by the blue band. The curves correspond to $\varepsilon=0.65$.
The data shown are: black open circle, PDG 2014~\cite{Agashe:2014kda};
blue circle, Olmos de Le{\'o}n et al~\cite{Olmos:2001};
green diamond, MIT-Bates (DR)~\cite{Bourgeois:2006js,Bourgeois:2011zz};
green open diamond, MIT-Bates (LEX)~\cite{Bourgeois:2006js,Bourgeois:2011zz};
purple solid square,
MAMI (DR)~\cite{Correa:thesis};
purple open square, MAMI (LEX)~\cite{Correa:thesis};
red solid triangle, MAMI1 (LEX)~\cite{Roche:2000ng};
red solid inverted triangle, MAMI1 (DR)~\cite{d'Hose:2006xz};
red open triangle, MAMI2 (LEX)~\cite{Janssens:2008qe}.
Some of the data points are shifted to the right 
in order to enhance their visibility; namely, Olmos de Le{\'o}n,
MIT-Bates (LEX),
MAMI LEX, MAMI1 DR
and MAMI2 LEX sets have the same values of $Q^2$ as PDG, MIT-Bates (DR), MAMI DR,
and MAMI1 LEX,
respectively.
}
\label{fig:response_function_PLL}
\end{figure}

The experiments aiming to measure the GPs are based on the low-energy expansion
of the $ep\ga$ process, \Eqref{LEX}, which results in the extraction of
the VCS response functions. Then, with some further assumptions on the size of spin GPs,
 taken usually from the
fixed-$t$ DR framework of Ref.~\cite{Drechsel:2002ar}, one obtains the two scalar GPs,
$\alpha_{E1}$ and
$\beta_{M1}$. We first consider our results at the level of the response functions, since
it provides a more direct comparison to experiment.

\begin{figure}[t]
\includegraphics[width=0.6\textwidth]{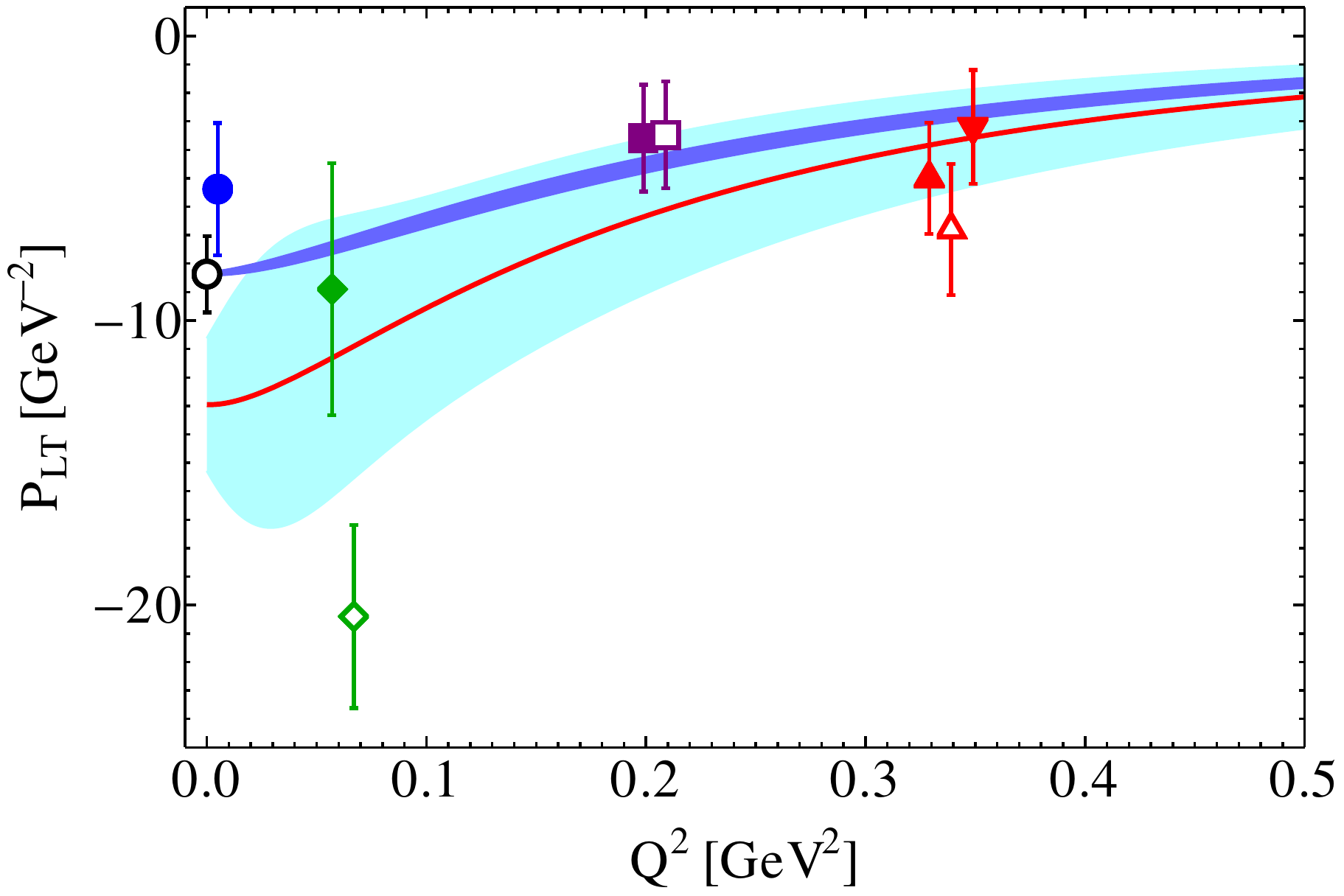}
\caption{VCS response function
$P_{LT}(Q^2)$.
Notation is as in Fig.~\ref{fig:response_function_PLL}.
}
\label{fig:response_function_PLT}
\end{figure}

\begin{figure}[ht]
\includegraphics[width=0.6\textwidth]{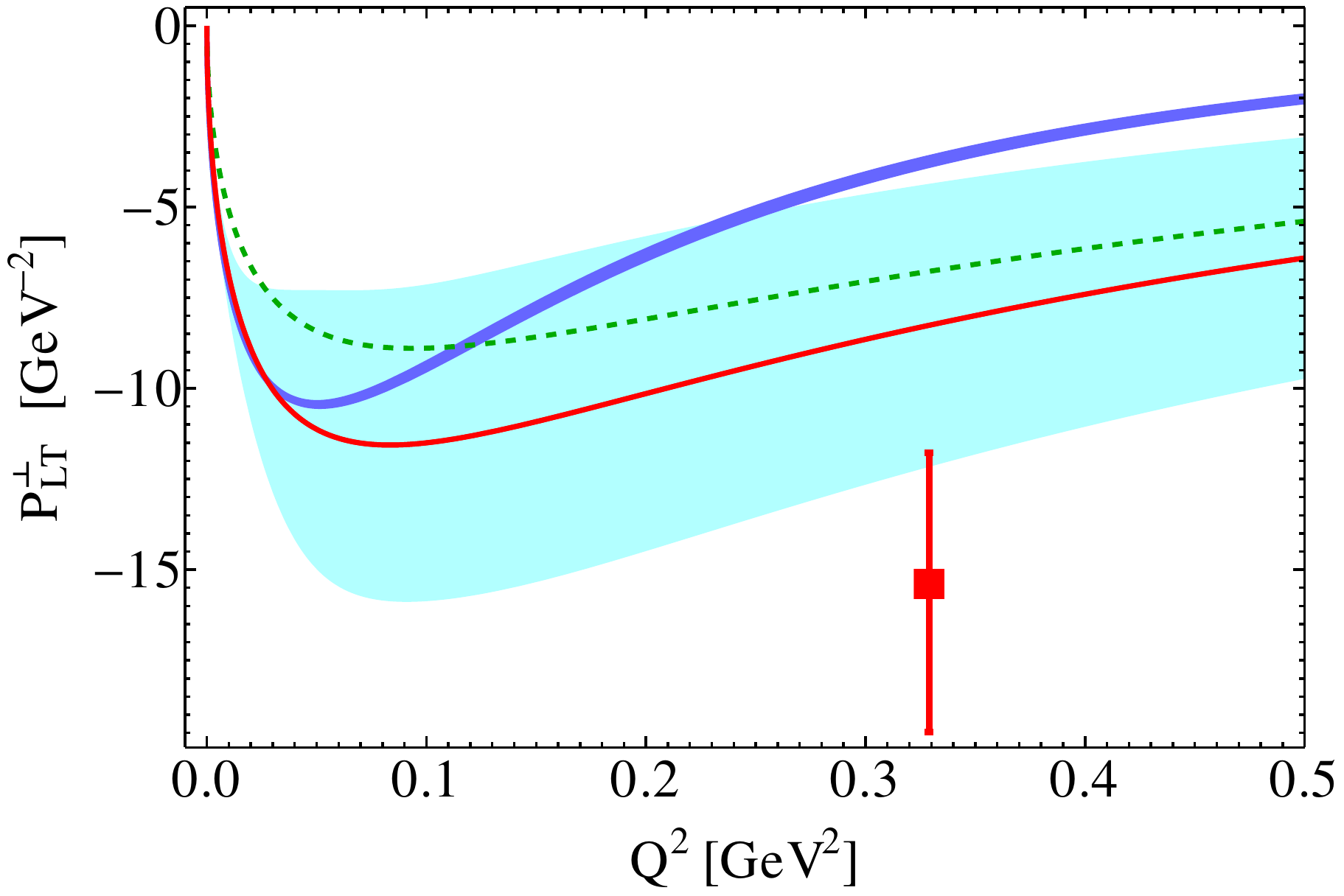}
\caption{VCS response function
$P_{LT}^{\perp}(Q^2)$.
Notation is as in Fig.~\ref{fig:response_function_PLL},
except from the data: red square, MAMI~\cite{Doria:2015dyx},
and the green dotted curve that shows the B$\chi$PT result
with only the contribution of $P_{LL}$ included, see Eq.~\eqref{eq:pltperp}.
}
\label{fig:response_function_PLTP}
\end{figure}

\begin{figure}[ht]
\includegraphics[width=0.6\textwidth]{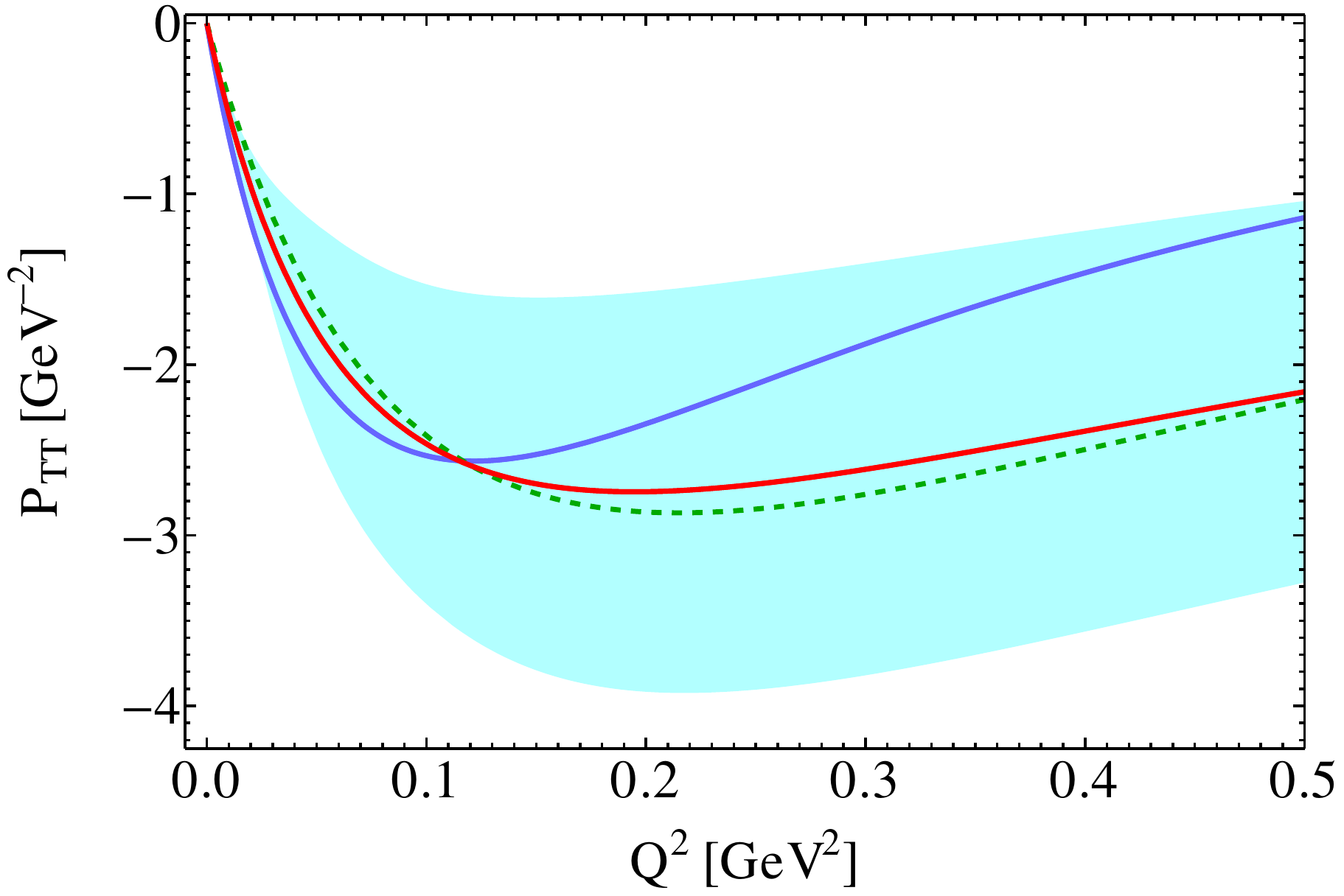}
\caption{VCS response function
$P_{TT}(Q^2)$.
Notation is as in Fig.~\ref{fig:response_function_PLL}, except from
the green dotted curve that shows the B$\chi$PT result with only
the contribution of $P^{(M1,M1)1}$ in Eq.~\eqref{eq:ptt}.
}
\label{fig:response_function_PTT}
\end{figure}

In Figs.~\ref{fig:response_function_PLL} to \ref{fig:response_function_PTT}, we show
our B$\chi$PT results (red solid line, with cyan band indicating the uncertainty estimate), 
compared with the fixed-$t$ DR calculation (blue bands), and experimental data where available.
In this calculation we used the
Bradford {\it et al.}~\cite{Bradford:2006yz} parametrization of nucleon form factors, as input in 
\Eqref{VCSresponse}. The bands of the DR results are obtained by varying
the dipole cut-offs
$\Lambda_{\alpha}$ and $\Lambda_\beta$ within the uncertainties given in
Sec.~\ref{sec:comparison:dr}.

The first two response functions,
$P_{LL}-P_{TT}/\varepsilon$ and $P_{LT}$ (Fig.~\ref{fig:response_function_PLL} 
and~\ref{fig:response_function_PLT}), are used to extract $\alpha_{E1}(Q^2)$ and
$\beta_{M1}(Q^2)$, respectively. Our results here are in good agreement with the data
as well as with the DR results.
The only place of disagreement is $P_{LT}(0) = -2M \beta_{M1}/\al_\mathrm{em}$, due to the larger value of the static magnetic
polarizability resulting in B$\chi$PT, as mentioned already in the previous section.


Apart from these two response functions
extracted from unpolarized measurements, there has been
a single low-$Q^2$ double-polarization experiment at MAMI~\cite{Doria:2015dyx}
extracting the response function $P_{LT}^\perp$ defined in \Eqref{pltperp}. 
This data point, together with theoretical curves, is shown in Fig.~\ref{fig:response_function_PLTP}.
This is perhaps the only place where one can see that the B$\chi$PT calculation is in a better agreement
with the data than the DR calculation. On the other hand, the slope at $Q^2=0$ is in a perfect agreement
between the two calculations.

This polarized observable can 
potentially provide an access to the spin GPs. For instance,
combining it with $P_{LL}-P_{TT}/\varepsilon$
one can extract the $P_{TT}$ response function, Fig.~\ref{fig:response_function_PTT}.
The latter is given entirely by the spin GPs. 
We note that in the $P_{TT}$ response function the large, and well known, $\pi^0$ $t$-channel pole 
contribution to several of the spin GPs drops out. 
We see from Fig.~\ref{fig:response_function_PTT} that the B$\chi$PT and DR 
results for $P_{TT}$ are again in reasonable agreement.

In  Figs.~\ref{fig:response_function_PLTP} and~\ref{fig:response_function_PTT}, we also show 
(by the green dashed curves) the dominant terms
in $P_{LT}^\perp $ and in $P_{TT}$. They are given by, respectively,
$P_{LL}$ and $P^{(M1,M1)1}$ terms in Eqs.~\eqref{eq:pltperp} and~\eqref{eq:ptt}. 
One thus sees, in particular, that $P^{(L1,M2)1}$, for which the 
B$\chi$PT and DR results differ sizeably at low $Q^2$,  
gives a very small contribution 
to $P_{TT}$.

\section{Concluding remarks}\label{sec:conclusions}

\begin{figure}[th]
\includegraphics[width=\textwidth]{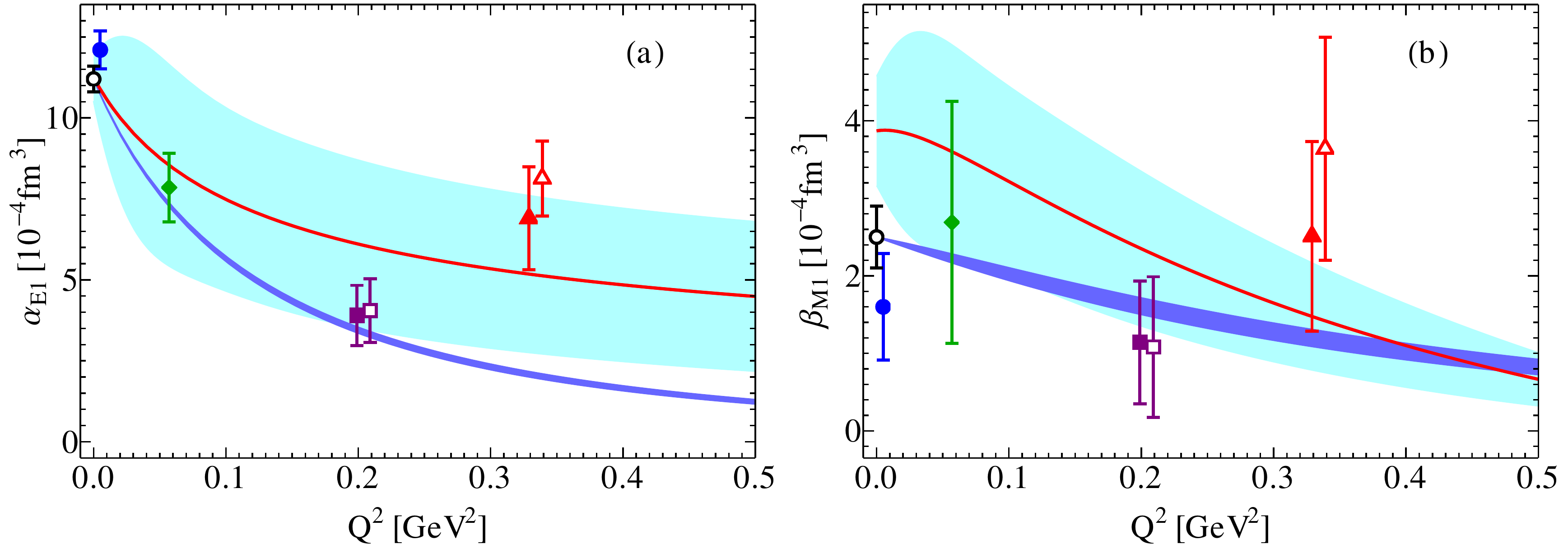}
\caption{Generalized scalar polarizabilities:
(a) $\alpha_{E1}(Q^2)$, (b) $\beta_{M1}(Q^2)$.
Description of curves and points is the same as in Fig.~\ref{fig:response_function_PLL}.
}
\label{fig:alpha_beta}
\end{figure}

The B$\chi$PT calculation of the nucleon GPs and VCS response functions,
presented here, is done to NLO in the $\de$-counting scheme. 
It shows a good description of the low-$Q$
data and mostly agrees with the results of the fixed-$t$ DR calculation of 
Pasquini {\it et al.}~\cite{Pasquini:2001yy}.
The results for the scalar GPs are summarized in Fig.~\ref{fig:alpha_beta},
where panel (a) shows the electric polarizability and panel (b)---the magnetic one.
The theoretical uncertainty of our calculation is sufficiently large to
agree with all the data, including the new~\cite{Correa:thesis}
and old~\cite{Roche:2000ng,Janssens:2008qe} MAMI data that
tend to disagree among themselves.
We can see that the DR curve agrees with the new MAMI data very well, 
while missing the older data, especially for $\al_{E1}$.
For $\beta_{M1}$, there is an interesting tension at low $Q$ between the DR and $\chi$PT
results. The available VCS data do not have the necessary precision
to resolve the discrepancy.

By making the heavy-baryon expansion we reproduce some of the previous 
HB$\chi$PT results and, similarly to what was observed in the calculation
of the real CS, we find that treating the leading chiral loops exactly
allows for a more natural accommodation of the Delta-resonance contribution,
which is especially large in the magnetic polarizability $\beta_{M1}$.

We would like to note that a newly approved experiment at Jefferson Lab~\cite{JLab_C12-15-001} which plans to 
measure the unpolarized GPs at $Q^2 = 0.3$~GeV$^2$ and $Q^2 = 0.75$~GeV$^2$ will be able to shed further light on the situation. Furthermore, comparing such data at the same $Q^2$ value taken at different values of $\varepsilon$ (corresponding with different beam energies) has the potential to separate off the response function $P_{TT}$ 
in~\Eqref{LEX}. This would allow one to experimentally access the dominant spin GP $P^{(M1,M1)1}$ for the first time and provide a strong test of the B$\chi$PT predictions presented in this work.   

Additionally, new data on the unpolarized response functions and GPs are
expected to arrive soon from MAMI. These data will complement the
$Q^2=0.2$~GeV${^2}$ points~\cite{Correa:thesis}.
In particular, expected are data at $Q^2=0.1$~GeV${^2}$ and $Q^2=0.45$~GeV${^2}$,
which is in the domain of applicability of B$\chi$PT. These data will also further test
the theoretical predictions.

One has to admit that the current theoretical uncertainty estimate
gives a rather sizeable error band, which should be improved upon. 
An $O(p^4)$ calculation
of GPs in B$\chi$PT that would include the remaining $\pi \Delta$ loops
that contribute at $O(p^3)$ in the high-momenta regime
and both the $\pi N$ and the $\pi\Delta$ $O(p^4)$ contributions
in this regime would allow one to significantly decrease the theoretical uncertainty.

\section*{Acknowledgements}
We thank H{\'e}l{\`e}ne~Fonvieille, Misha~Gorchtein, Chungwen~Kao, and Barbara~Pasquini 
for stimulating discussions and helpful communications.
This work was supported by the Deutsche Forschungsgemeinschaft (DFG) through the
Collaborative Research Center ``The Low-Energy Frontier of the Standard Model'' (SFB 1044)
and the Cluster of Excellence PRISMA. 
V.~L.\ acknowledges partial support of this work by the Moscow Engineering
Physics Institute Academic Excellence Project (Contract No.\ 02.a03.21.0005).
We acknowledge the use of FORM~\cite{Vermaseren:2000nd} in the calculations
and of JaxoDraw~\cite{Binosi:2008ig} in preparation of the figures.

\appendix
\small

\section{Tensor decomposition of the VCS amplitude}
\label{sec:appendix:tensors}

In this section we give the details of the tensor decomposition of the VCS
amplitude. The basis used by us is $\rho_i$, $i=1,\dots, 12$, introduced in
Ref.~\cite{Drechsel:1997xv}. Its decomposition in terms of Tarrach's
$T_{1},\dots,T_{34}$ (which are given below) reads
\begin{align}
\rho_1 =& -q\cdot q' T_1 + T_3\,,\displaybreak[0]\nonumber \\
\rho_2 =& -4M^2\xi^2 T_1-4q\cdot q' T_6+4M\xi T_7\,,\displaybreak[0]\nonumber\\
\rho_3 =& -2M\xi Q^2 T_1 - M\xi (T_4+T_5)+Q^2(T_7-T_8)+q\cdot q'(T_9-T_{10})\,,\displaybreak[0]\nonumber\\
\rho_4 =& 8 T_{16}-4M\xi T_{21}+M\xi T_{34}\,,\displaybreak[0]\nonumber\\
\rho_5 =& \frac{1}{2}(T_{19}-T_{20})-\frac{Q^2}{2}T_{22}-\frac{M\xi}{2}(T_{23}+T_{24})-\frac{Q^2}{8}T_{34}\,,\displaybreak[0]\nonumber\\
\rho_6 =& -8q\cdot q' T_6+4M\xi T_7 +4M q\cdot q' T_{21}-4M^2\xi T_{25}-2M\xi(T_{32}+q\cdot q'T_{33})+M q\cdot q' T_{34}\,,\displaybreak[0]\nonumber\\
\rho_7 =& T_{18}-q\cdot q' T_{22}+M\xi T_{26}\,,\displaybreak[0]\nonumber\\
\rho_8 =& \frac{M\xi}{2} (T_4+T_5)+\frac{Q^2}{2}T_8 -\frac{q\cdot q'}{2}(T_9-T_{10})-\frac{M}{2}(T_{14}+T_{15})+\frac{Mq\cdot q'}{2}(T_{23}+T_{24})+\frac{MQ^2}{2}T_{26}\nonumber\\
&+\frac{Q^2}{4}(T_{32}+q\cdot q' T_{33})\,,\displaybreak[0]\nonumber\\
\rho_9 =& 2M\xi T_8-2Mq \cdot q' T_{22}+2M^2\xi T_{26}-q\cdot q' T_{27}+M\xi T_{31}\,,\displaybreak[0]\nonumber\\
\rho_{10} =& 2T_7 +4M T_{11}-2M T_{25}-4M\xi T_1+(T_{32}+q\cdot q'T_{33}) \,,\displaybreak[0]\nonumber\\
\rho_{11} =& 4T_{17}-4M\xi T_{25}+q\cdot q' T_{34}\,,\displaybreak[0]\nonumber\\
\rho_{12} =& 2Q^2 T_6 +M\xi(T_9-T_{10})-MQ^2(T_{21}+T_{22})-M^2\xi(T_{23}+T_{24})-\frac{Q^2}{2}(T_{27}-M\xi T_{33})\nonumber\\
&-\frac{M\xi}{2}(T_{29}+T_{30})-\frac{MQ^2}{4}T_{34}\,.
\end{align}
These tensors correspond to the following combinations of Tarrach's
$\tau_i$ (with $q'^2$ set to zero in the latter):
\begin{align}
\rho_1 &= -\tau_1, & \rho_2 &= -4\tau_3, & \rho_3 &=\tau_4-\tau_5, & \rho_4 &=\tau_7,\nonumber \\
\rho_5 &=\frac{1}{2}(\tau_8-\tau_9),& \rho_6 &=\tau_{10}, &\rho_7 &=\tau_{11}, & \rho_8 &=\frac{1}{2}(\tau_{12}+\tau_{13}),\\
\rho_9 &=\tau_{14},& \rho_{10} &=\tau_{17}, & \rho_{11} &=\tau_{18},& \rho_{12}&=\frac{1}{2}(\tau_{20}+\tau_{21})\nonumber\,.
\end{align}
All tensors apart from $\rho_2$, $\rho_3$, and $\rho_6$ have unique
structures that allow for unambiguous identification, e.g., the combination
$T_{29}+T_{30}$ enters only $\rho_{12}$, $T_{17}$ enters only $\rho_{11}$, and so on.
After the tensors $\rho_i$, $i\neq 2,3,6$, have been identified,
the remaining tensors can be identified as well.
Since this basis is explicitly gauge invariant, all the terms that
are not proportional to any of $\rho_i$ have to vanish when one decomposes
a gauge invariant amplitude, e.g., summing up a gauge invariant subset of Feynman graphs.

The tensors $T_{1},\dots, T_{34}$ introduced by Tarrach~\cite{Tarrach:1975tu}
in order to decompose the CS amplitude in the most general case,
i.e., when both $q^2$ and $q'^2$ are non-zero, are given below;
these structures are understood to be contracted
with $\epsilon^\nu$ and $\epsilon^{*\mu}$,
the incoming and the outgoing photons' polarization vectors.
\begin{align}
T_1=g_{\mu\nu} \,,&\qquad T_{17}=\left(P_\nu q_\mu + P_\mu q'_\nu\right)\slashed{K}\,,\displaybreak[0]\nonumber\\
T_2=q_\nu q'_\mu \,,&\qquad T_{18} = \left(P_\nu q_\mu -P_\mu q'_\nu\right)\slashed{K}\,,\displaybreak[0]\nonumber\\
T_3=q'_\nu q_\mu \,,&\qquad T_{19} = \left(P_\nu q'_\mu+P_\mu q_\nu\right)\slashed{K}\,,\displaybreak[0]\nonumber\\
T_4=q_\nu q_\mu + q'_\nu q'_\mu \,,&\qquad T_{20} = \left(P_\nu q'_\mu-P_\mu q_\nu\right)\slashed{K}\,,\displaybreak[0]\nonumber\\
T_5=q_\nu q_\mu -q'_\nu q'_\mu \,,&\qquad T_{21}=P_\nu \gamma_\mu+P_\mu\gamma_\nu\,,\displaybreak[0]\nonumber\\
T_6=P_\nu P_\mu \,,&\qquad T_{22}=P_\nu \gamma_\mu-P_\mu\gamma_\nu\,,\displaybreak[0]\nonumber\\
T_7=P_\nu q_\mu+P_\mu q'_\nu \,,&\qquad T_{23}=q_\nu\gamma_\mu+q'_\mu\gamma_\nu\,,\displaybreak[0]\nonumber\\
T_8=P_\nu q_\mu - P_\mu q'_\nu \,,&\qquad T_{24}=q_\nu\gamma_\mu-q'_\mu\gamma_\nu\,,\displaybreak[0]\nonumber\\
T_9=P_\nu q'_\mu+P_\mu q_\nu \,,&\qquad T_{25}=q'_\nu\gamma_\mu+q_\mu\gamma_\nu\,,\displaybreak[0]\nonumber\\
T_{10}=P_\nu q'_\mu-P_\mu q_\nu \,,&\qquad T_{26}=q'_\nu \gamma_\mu-q_\mu \gamma_\nu\,,\displaybreak[0]\\
T_{11}=g_{\mu\nu}\slashed{K} \,,&\qquad T_{27}=2\left(P_\nu \gamma_{\mu\lambda}K^\lambda +P_\mu\gamma_{\nu\lambda}K^\lambda\right)\,,\displaybreak[0]\nonumber\\
T_{12}=q_\nu q'_\mu\slashed{K} \,,&\qquad T_{28}=2\left(P_\nu \gamma_{\mu\lambda}K^\lambda -P_\mu\gamma_{\nu\lambda}K^\lambda\right)\,,\displaybreak[0]\nonumber\\
T_{13}=q'_\nu q_\mu \slashed{K} \,,&\qquad T_{29}=2\left(q_\nu\gamma_{\mu\lambda}K^\lambda+q'_\mu\gamma_{\nu\lambda}K^\lambda\right)\,,\displaybreak[0]\nonumber\\
T_{14}=(q_\nu q_\mu+q'_\nu q'_\mu)\slashed{K} \,,&\qquad T_{30}=2\left(q_\nu\gamma_{\mu\lambda}K^\lambda-q'_\mu\gamma_{\nu\lambda}K^\lambda\right)\,,\displaybreak[0]\nonumber\\
T_{15}=(q_\nu q_\mu-q'_\nu q'_\mu)\slashed{K} \,,&\qquad T_{31}=2\left(q'_\nu\gamma_{\mu\lambda}K^\lambda+q_\mu\gamma_{\nu\lambda}K^\lambda\right)\,,\displaybreak[0]\nonumber\\
T_{16}=P_\nu P_\mu \slashed{K} \,,&\qquad T_{32}=2\left(q'_\nu\gamma_{\mu\lambda}K^\lambda-q_\mu\gamma_{\nu\lambda}K^\lambda\right)\,,\displaybreak[0]\nonumber\\
   &\qquad T_{33} = 2\gamma_{\nu\mu}\,,\displaybreak[0]\nonumber\\
   &\qquad T_{34} = 2\left\{\gamma_{\nu\mu},\slashed{K}\right\}=4\gamma_{\nu\mu\lambda}K^\lambda\,.\nonumber
\end{align}
Here, $P=\half (p+p')$, $K=\half(q+q')$, $\gamma^{\mu\nu}=\half[\gamma^\mu,\gamma^\nu]$ and
$\gamma^{\mu\nu\lambda}=\half\{\gamma^{\mu\nu},\gamma^\lambda\}$.
The following relations hold between these tensors that allow one to exclude two
of them (the usual choice being $T_{13}$ and $T_{28}$):
\begin{equation}
\begin{split}
2(T_{17}-T_{19})-(q^2-q'^2)T_{22}+2P\cdot K(T_{23}-T_{25})&-2MT_{28}-2MP\cdot K\, T_{32}\\
&+\left(M^2+\frac{q.q'}{4}-\frac{q^2+q'^2}{4}\right)T_{34}=0\,,
\end{split}
\end{equation}
\begin{equation}
\begin{split}
&P\cdot K(T_2-T_3)+\frac{1}{4}(q^2+q'^2+2q\cdot q')(T_7-T_9)-\frac{q^2-q'^2}{4}(T_8+T_{10})-M(T_{12}-T_{13})\\
&+\frac{M}{4}(q^2+q'^2+2q\cdot q')(T_{23}-T_{25})-M\frac{q^2-q'^2}{4}(T_{24}+T_{26})-P\cdot K\, T_{28}+\frac{q^2-q'^2}{8}(T_{29}-T_{31})\\
&-\frac{1}{8}(q^2+q'^2-2q\cdot q')(T_{30}+T_{32})-\left[\left(P\cdot K\right)^2+\frac{1}{4}\left(q^2q'^2-(q\cdot q')^2\right)\right]T_{33}+\frac{M}{2}P\cdot K\, T_{34}=0\,.
\end{split}
\end{equation}
Taking into account the fact that for the (real) final photon $\epsilon'\cdot q'=0$, one can obtain the following useful identities:
\begin{equation}
T_2=0,\quad T_4=T_5,\quad T_9=-T_{10},\quad T_{12}=0,\quad T_{14}=T_{15},\quad T_{19}=-T_{20}, \quad T_{23}=T_{24},\quad T_{29}=T_{30}\,.
\end{equation}

\section{Invariant amplitudes}\label{sec:appendix:amplitudes}

Here we provide expressions for the linear combinations of invariant
amplitudes $\bar{A}_i(Q^2)=\bar{A}_i(0,-Q^2,-Q^2)$ that contribute to the generalized polarizabilities, see Eqs.~\eqref{eq:PL1L10}-\eqref{eq:PL1M21}:
\begin{align}
g_1=\bar{A}_1,\ g_2=\bar{A}_2,\ g_3=\bar{A}_5,\ g_4=\bar{A}_5+\bar{A}_7+4\bar{A}_{11},
\ g_5=2\bar{A}_6+\bar{A}_9,\ g_6=\bar{A}_{12}\,.
\end{align}
The results are given for the $\pi$N loop and Delta pole contributions; for the $\pi\Delta$ loop results, see supplementary
material to this article.

\subsection{$\pi N$ loops}

Here, $D_1(x,y) = \left[\tau (1 - x)^2 (1-4y^2) + D_N(x)\right]^{-1}$,
$D_2(x,y) = \left[\tau x^2(1-4y^2)  + D_N(x)\right]^{-1}$, 
$D_0(x,y) = \left[D_N(x)\right]^{-1}$, and $D_\pi(x,y)=\left[4\tau (1 - x) x + \mu^2\right]^{-1}$. In turn,
$D_N(x)=x^2+\mu^2(1-x)$. The amplitudes
are expressed as integrals over the Feynman parameters as follows:
\begin{equation}
g_i(Q^2)=\frac{g_A^2}{8\pi^2 f_\pi^2 M^{n_i}}\int\limits_0^1\mathrm{d}x\int\limits_{-1/2}^{1/2}\mathrm{d}y\, \phi_i(\tau,x,y)\,,
\end{equation}
where $\phi_i$ are given below, $g_A=1.27$ and $f_\pi=92.21$~MeV are the axial coupling
constant and the pion decay constant,  and $n_i=1,3,2,2,3,3$ for $i=1,2,3,4,5,6$
account for the correct dimensions of the respective $g_i$.
\subsubsection{Proton}
\begin{align}
\phi_1=&-\frac{1}{2} D_1^2(x,y) (x-1)^3 x \left(-4 y ^2+2 x+1\right)-\frac{1}{4} D_2^2(x,y)
x^4 \left(4 y ^2+2 x-1\right)\nonumber\\
&
-\frac{1}{2} D_0^2(x,y) x^3 \left(3 x^2-5 x+2\right)-2 D_\pi(x,y) (x-1)^2 x\,,
 \displaybreak[0]
\\
\phi_2=&4 D_1^4(x,y) y ^2 \left(4 y ^2-1\right) \tau (x-1)^6 x^3\nonumber\\
&-\frac{1}{3} D_1^3(x,y)(x-1)^4 x \left(4y ^2 \left(4 y ^2-1\right) \tau+x^2 \left(8 y ^2 \tau-1\right)-4y ^2 \left(4 y ^2+1\right)\tau x\right)\nonumber\\
&+\frac{1}{24} D_1^2(x,y) (x-1)^3 x
   \left(\left(36 y ^2-7\right) x+4\right)\nonumber\\
   &+   2 D_2^4(x,y) y ^2 \left(4 y ^2-1\right) \tau (x-1)^2 x^7\nonumber\\
   &-\frac{1}{12} D_2^3(x,y) (x-1) x^5 \left(\tau \left(-48 y ^4+2 \left(36 y ^4-5 y ^2+1\right) x-1\right)+2 \left(3 y ^2
   (x-2)-1\right)\right)\nonumber\\
   &+
   \frac{1}{16}D_2^2(x,y) x^3 \left(-4 y ^2+\left(4 y ^2+5\right) x^2-8 x+3\right)
   \nonumber\\
   & -\frac{3}{2} D_0^4(x,y) (x-1)^2 x^4 \left(\mu ^2+\left(\mu ^2-1\right) x^2-2 \mu ^2 x\right)\nonumber\\
   &+ \frac{1}{12} D_0^3(x,y) (x-1) x^2 \left(-18 \mu ^2+10
   x^4+9 \left(\mu ^2-3\right) x^3+\left(26-36 \mu ^2\right) x^2+45 \mu ^2 x\right)
   \nonumber\\
   &-\frac{1}{6}D_0^2(x,y) (x-2) (x-1)^2
   \,, \displaybreak[0]\\
\phi_3=&\frac{1}{2} D_1^2(x,y) (x-1)^2 x \left(8 y ^2+\left(8 y ^2+1\right) x^2-16 y ^2 x\right)+\frac{1}{4} D_2^2(x,y) x^4 \left(8 y ^2 x+x-1\right)
\nonumber\\
&+\frac{1}{4} D_0^2(x,y) x^3 \left(3 x^2-5 x+2\right)\,, \displaybreak[0]\\
\phi_4 =&\frac{1}{2} D_1^2(x,y) (x-1)^3 x \left(4 y ^2 (4 x-3)+1\right)+\frac{1}{4} D_2^2(x,y) x^4 \left(4 y ^2 (4 x-3)-1\right)\,, \displaybreak[0]\\
\phi_5  =&-2 D_1^3(x,y) y ^2 \left(4 y ^2-1\right) \tau (x-1)^5 x^2-\frac{1}{4} D_1^2(x,y) (x-1)^3 x \left(-4 y ^2+\left(12 y ^2-1\right) x+1\right)\nonumber\\
&+
\frac{1}{12} D_2^3(x,y) (x-1) x^5 \left(-12 y ^2+\left(4 y ^2-1\right) \tau
   \left(-8 y ^2+2 \left(6 y ^2-1\right) x+1\right)+2 x-1\right)\nonumber\\
&
-\frac{1}{48} D_2^2(x,y) x^3 \left(-12 y ^2+16 x^2+\left(12 y ^2-25\right) x+9\right)\nonumber\\
&+\frac{3}{2} D_0^4(x,y) (x-1)^2 x^4 \left(\mu ^2+\left(\mu ^2-1\right) x^2-2 \mu ^2
   x\right)\nonumber\\
&-\frac{1}{12} D_0^3(x,y) (x-1) x^2 \left(-18 \mu ^2+10 x^4+9 \left(\mu ^2-3\right) x^3+\left(26-36 \mu ^2\right) x^2+45 \mu ^2 x\right)\nonumber\\
&+\frac{1}{12} D_0^2(x,y) (x-1)^2 x^2 (5 x-6)\,, \displaybreak[0]\\
\phi_6  =&-4 D_1^3(x,y) y ^2 (x-1)^4 x^3-2 D_2^3(x,y) y ^2 (x-1) x^6\,.
\end{align}

\subsubsection{Neutron}
\begin{align}
\phi_1=&\frac{1}{2} D_1^2(x,y) (x-1)^2 x \left(-4 y ^2+4 y ^2 x+x+1\right)-\frac{1}{2}D_2^2(x,y) \left(4 y ^2+1\right) x^4
\nonumber\\
&-2 D_\pi(x,y) (x-1)^2 x\,,
 \displaybreak[0]\\
\phi_2=&-4D_1^4(x,y) y ^2 \left(4 y ^2-1\right) \tau (x-1)^5 x^3\nonumber\\
&+\frac{1}{12} D_1^3(x,y) (x-1)^2 x \nonumber\\
&\hspace{1cm}\times 
\left(4 y ^2 \tau \left(x^2+x-2\right)^2-x^2 \left(\tau (x-1)^2+x (x+2)-4\right)+64 y ^4 \tau (x-1)^3\right)\nonumber\\
&+\frac{1}{24} D_1^2(x,y) (x-1)^2 x (x (4 x-1)-4)\nonumber\\
&+4D_2^4(x,y) y ^2 \left(4 y ^2-1\right) \tau (x-1)^2 x^6\nonumber\\
&-\frac{1}{12} D_2^3(x,y) (x-1) x^4\nonumber\\
&\hspace{1cm}\times 
\left(4 y ^2 \left(\tau \left(2 x^2+x-8\right)+3 (x-2) x\right)-48 y ^4 \tau (x-2) x+x \left(\tau (x+1)+x-3\right)-2\right)\nonumber\\
&+\frac{1}{24}D_2^2(x,y) (x-1) x^2 \left(4 x^2+24 y ^2 (2 x-1) x+x+1\right)\,, \displaybreak[0]\\
\phi_3=&-\frac{1}{2} D_1^2(x,y) (x-1) x \left(8 y ^2+\left(8 y ^2+1\right) x^2-16 y ^2 x\right)+\frac{1}{2} D_2^2(x,y) x^3 \left(8 y ^2 x+x-1\right) \,, \displaybreak[0]\\
\phi_4 =&\frac{1}{2} D_1^2(x,y) (x-1)^3 x \left(4 y ^2 (4 x-3)+1\right)+\frac{1}{4} D_2^2(x,y) x^4 \left(4 y ^2 (4 x-3)-1\right)\,, \displaybreak[0]\\
\phi_5  =&-\frac{1}{6} D_1^3(x,y) (x-1)^2 x^3 \left(\left(4 y ^2-1\right) \tau (x-1)^2- (x-2) x\right)\nonumber\\
&
-\frac{1}{12} D_1^2(x,y) (x-1)^2 x \left(12 y ^2+4 x^2-3 \left(4 y ^2+1\right) x-3\right)\nonumber\\
&
-\frac{1}{12} D_2^3(x,y) (x-1) x^5 \left(\left(4 y^2-1\right) \tau \left(-20 y ^2+12 y ^2 x+x+1\right)+36 y ^2-12 y ^2 x-x-1\right)
\nonumber\\
&
-\frac{1}{24} D_2^2(x,y) (x-1) x^3 \left(-36 y ^2+\left(48 y ^2+4\right) x+3\right)\,, \displaybreak[0]\\
\phi_6  =&4 D_1^3(x,y) y ^2 (x-1)^3 x^3-4 D_2^3(x,y)y ^2 (x-1) x^5\,.
\end{align}

\subsection{Delta pole}
Here, $g_E$ and $g_M$ are the electric and magnetic $\gamma N\Delta$
couplings~\cite{Pascalutsa:2002pi}, $M_+=M_\Delta+M$, and 
$$f_M=\frac{g_M}{\big[1+\left(Q/\Lambda\right)^2\big]^2}$$
is the magnetic $\gamma N\Delta$ coupling modified by the dipole form factor,
with $\Lambda^2=0.71$~GeV${^2}$.

\begin{align}
g_1&=\frac{ f_M \left[2 g_M\, M_+ \left(4 M^2+Q^2\right)-g_E\, Q^2\varDelta   \right]+g_E\, Q^2 (g_M\, M_+-2g_E\, \varDelta )}{4M^2M_+^3 \varDelta  }\,,
\displaybreak[0]\\
g_2&=-\frac{f_M\, g_M\, M_+-g_E^2\,\varDelta  }{2  M^2 M_+^3\varDelta}\,,
\displaybreak[0]\\
g_3&=\frac{f_M \left(8 M^2+Q^2\right) (g_M M_+ -2 g_E\,\varDelta  )+g_E \left[2 g_M\, M_+ Q^2- g_E \,\varDelta \left(8 M^2+Q^2\right)\right]}{8  M^3 M_+^3\varDelta}\,,
\displaybreak[0]\\
g_4 &=\frac{f_M\, g_M\, M_+-2 f_M\, g_E\,\varDelta  -2 g_E^2\,\varDelta  +3 g_E\, g_M\, M_+}{M M_+^3\varDelta }\,,
\displaybreak[0]\\
g_5 & =-\frac{f_M\, g_M\, M_+-3f_M\, g_E\, \varDelta  -g_E^2\,\varDelta  +3 g_E\, g_M\, M_+}{4  M^2 M_+^3\varDelta}\,,
\displaybreak[0]\\
g_6 & = -\frac{f_M\, g_M\, M_+-2f_M\, g_E\, \varDelta  -g_E^2\,\varDelta  +2 g_E\, g_M\, M_+}{4 M^2M_+^3 \varDelta }\,.
\end{align}

\end{document}